\documentclass[prd,twocolumn]{revtex4}
\usepackage{amsmath}
\usepackage{graphicx}
\usepackage{amsfonts}
\usepackage{amssymb}
\usepackage{bm}
\usepackage{epsfig}
\usepackage{amsmath}
\newcommand{\be}{\begin{equation}}
\newcommand{\ee}{\end{equation}}
\newcommand{\ba}{\begin{eqnarray}}
\newcommand{\ea}{\end{eqnarray}}
\newcommand{\bml}{\begin{mathletters}}
\newcommand{\eml}{\end{mathletters}}
\newcommand{\bes}{\begin{subequations}}
\newcommand{\ees}{\end{subequations}}
\newcommand{\ord}{{\cal O}}
\newcommand{\bi}{\begin{itemize}}
\newcommand{\ei}{\end{itemize}}
\newcommand{\gev}{~{\rm GeV}}
\newcommand{\tev}{~{\rm TeV}}

\begin{document}
\title{$E_6$ unification in a model of dark energy and dark matter}
\author{P.Q. Hung}
\email[]{pqh@virginia.edu}
\affiliation{Dept. of Physics, University of Virginia, \\
382 McCormick Road, P. O. Box 400714, Charlottesville, Virginia 22904-4714,
USA}
\author{Paola Mosconi}
\email[]{pm8u@virginia.edu}
\affiliation{Dept. of Physics, University of Virginia, \\
382 McCormick Road, P. O. Box 400714, Charlottesville, Virginia 22904-4714,
USA}

\date{\today}
\begin{abstract}
A model of dark energy and dark matter was proposed earlier by one
of us (PQH) which involved an unbroken gauge group $SU(2)_Z$ whose
coupling $\alpha_Z \equiv g_Z^2/4\,\pi\sim O(1)$ at a scale
$\Lambda_Z \sim 3 \times 10^{-3}\,eV$ starting from a value within
the range of the Standard Model (SM) couplings at a high energy
scale $\sim 10^{16}\,GeV$. In that model, the universe is assumed to
be presently trapped in a false vacuum with an energy density $\sim
\Lambda_Z^4$. In this paper, we present a scenario in which
$SU(2)_Z$ is unified with the SM through several steps: $E_6
\rightarrow SU(2)_Z \otimes SU(6) \rightarrow SU(2)_Z \otimes
SU(3)_c \otimes SU(3)_L \otimes U(1) \rightarrow SU(2)_Z \otimes
SU(3)_c \otimes SU(2)_L \otimes U(1)_Y$. This unification provides a
rationale for why the value of the $SU(2)_Z$ coupling is within the
range of the SM couplings at high energies. The particle content and
the route of symmetry breaking in this model is very different from
the usual $E_6$ unification encountered in the literature. Several
implications, in addition to the dark energy, include the existence
of heavy mirror particles which could be searched for at future
colliders such as the LHC.
\end{abstract}
\maketitle

\section{Introduction}

The origin of the so-called dark energy responsible for the acceleration
of the present universe is one of the deepest problems for at least
the next several decades. In the quest for an understanding of that
origin, there are several approaches that are worth mentioning. One is
the improvement of cosmological and astrophysical observations which
have become increasingly precise. This will allow us to study, among
other things, the equation of state of the dark energy
$p= w\, \rho$ as a function
of redshift, for example.

Somewhat coupled to the observational effort is the various
theoretical enterprizes aimed at trying to make sense out of the
discovery of the accelerating universe. In particular, several
models were constructed: quintessence, modified gravity, etc...
However, the most recent determination of the equation of state $w
\sim -1$ appears to be consistent
with the present universe which is dominated by a cosmological
constant and cold dark matter, the so-called $\Lambda CDM$ scenario
\cite{DE}.
Although it is probably too early to decide which scenario is the
most plausible one, it might be interesting to build a dynamical
model which can practically mimic the $\Lambda CDM$ scenario at the
present time.

In \cite{hung1}, \cite{hung2}, such a model was constructed in which
it was proposed that the present universe is trapped in a false
vacuum characterized by an energy density $\rho_V \sim (3 \times
10^{-3} \,eV)^4$. A detailed description of the origin of the
potential which describes this false vacuum was given in
\cite{hung2}. In a nutshell, this is a potential of a
pseudo-Nambu-Goldstone (PNG) boson $a_Z$ (the imaginary part of a
singlet scalar field $\phi_Z$) coming from a spontaneously broken
global symmetry $U(1)_{A}^{(Z)}$ present in the model. It is induced
by instantons of a new gauge group $SU(2)_Z$ which grows strong at a
scale $\sim 3 \times 10^{-3} \,eV$ \cite{goldberg}. 
In \cite{hung2}, it was shown
how $\alpha_Z = g_Z^2/4\pi$ becomes order of unity at $ \Lambda_Z
\sim 3 \times 10^{-3} \,eV$ starting from an initial value at $M
\sim 10^{16}\,GeV$ of the order of the Standard Model (SM) couplings
at a similar scale. It was suggested in \cite{hung2} that this
behavior could arise from some form of unification of $SU(2)_Z$ with
the SM into the gauge group $E_6$ with the following suggested
symmetry breaking pattern: $E_6 \rightarrow SU(2)_Z \otimes SU(6)
\rightarrow SU(2)_Z \otimes SU(3)_c \otimes SU(3)_L \otimes U(1)_6
\rightarrow SU(2)_Z \otimes SU(3)_c \otimes SU(2)_L \otimes U(1)_Y$.
(Notice that the group $SU(3)_L \otimes U(1)$ was discussed
by several authors in the early nineties and an incomplete
list of references is given in \cite{331}.)
In this sense, the emergence of a new (unbroken) strong gauge group
$SU(2)_Z$ from the Grand Unified Theory (GUT) group $E_6$ and its
subsequent evolution is well motivated. In this paper, we will
present a detailed discussion of this unification scenario.

In passing, it is worth mentioning that the model presented in \cite{hung2}
also contained several interesting implications such as the presence of
cold dark matter candidates which are the fermions transforming as
adjoints of $SU(2)_Z$ and are SM singlets; and a new mechanism of
leptogenesis \cite{hung3} involving the decay of a ``messenger'' scalar field
${\tilde{\bm{\varphi}}}^{(Z)}$ which
carries the quantum numbers of both $SU(2)_Z$ and $SU(2)_L \otimes U(1)_Y$.
Another cosmological implication is the possibility that the real part
of the aforementioned singlet field $\phi_Z$, namely $\sigma_Z$, can
play the role of the inflaton in a ``low scale'' inflationary scenario
\cite{barcelona}. There it is seen that the role of the messenger field
is crucial both in generating SM particles at the end of inflation
through thermalization and in the subsequent leptogenesis (above the
electroweak scale) which converts a net SM lepton number into a net
baryon number through the electroweak sphaleron. In addition, it was
emphasized in \cite{hung2} and \cite{hung3} that one actually look
for the messenger field- called the
``lepton number progenitor'' in  \cite{hung3}-itself at future 
colliders such as the LHC (or
even the proposed ILC) since its mass is constrained by the
leptogenesis scenario to be less than $1\,TeV$ \cite{hung3}.

The plan of the paper will be as follows. First, we will list the
particle content of the model in \cite{hung1}, \cite{hung2} and show
how it fits into representations of $E_6$. Next, we discuss the
pattern of symmetry breaking of the model, including discussions on
fermion and scalar masses. We then follow with an analysis of the
renormalization-group evolution of the various gauge couplings. In
particular we will compute the two relevant unification scales:
$M_6$ where $SU(6)$ is broken down and $M_{GUT}$ where $E_6$ is
broken down. It is the value of $M_6$ which is most relevant to the
proton lifetime as we shall see. We will end with some brief comments
in the conclusion
on the experimental implications of our model, in
particular the possible detection of various particles which are
present in the model such as mirror fermions.

\section{Fitting the particle content of the dark energy model
into $E_6$}

We now list the particle content of the dark energy model of
\cite{hung1}, \cite{hung2}. (The latter reference contains extensive
details of that model.) As we shall see below, the effective gauge
group just above the electroweak breaking scale is $SU(2)_Z \otimes
SU(3)_c \otimes SU(2)_L \otimes U(1)_Y$. We concentrate primarily on
the content of particles which are {\em non-singlet} under $SU(2)_Z$
and on a complex scalar field which is singlet under both sectors.
SM particles are all $SU(2)_Z$ {\em singlets} and we shall not need
to list them explicitly here.

\subsection{Particle content (other than the SM one) of the dark energy model
\cite{hung2}}

I) $SU(2)_Z$-non-singlet fermions:
Under $SU(2)_Z \otimes SU(3)_c \otimes SU(2)_L \otimes U(1)_Y$,
these transform as
\be
\label{psiZ}
\psi_{L,R}^{(Z),i}=(3,1) \,,
\ee
where $i=1,2$. The fermions $\psi_{i}^{(Z)}$ were shown to have the appropriate
masses  (O($100-200\,GeV$)) and annihilation cross sections (typical of a weak
cross section) to be considered as candidates for
the WIMP cold dark matter \cite{hung1}, \cite{hung2}.

II) $SU(2)_Z$-non-singlet scalars:
\be
\label{messenger}
{\tilde{\bm{\varphi}}}_{i}^{(Z)} =({\tilde{\bm{\varphi}}}_{i}^{(Z),0},
{\tilde{\bm{\varphi}}}^{(Z),-}_{i})=
(3, 1,2,Y_{\tilde{\varphi}}=-1) \,,
\ee
where $i=1,2$. These are the so-called messenger fields since they carry quantum
numbers of both the SM and $SU(2)_Z$ sectors. (One of them,
${\tilde{\bm{\varphi}}}_{2}^{(Z)}$, is constrained in \cite{hung2}
and \cite{hung3}
to be much heavier than the other ${\tilde{\bm{\varphi}}}_{1}^{(Z)}$ whose
mass is constrained to be around $\sim 300-1000\,GeV$.)
As such, they have Yukawa
couplings with SM leptons and $\psi_{i}^{(Z)}$. It was shown in \cite{hung3}
that it is this kind of SM-lepton-number-violating coupling which
can give rise to a new mechanism of leptogenesis through a CP-violating
decay of the lightest of the two messenger fields, namely
${\tilde{\bm{\varphi}}}_{1}^{(Z)}$.

III) $SU(2)_Z \otimes SU(3)_c \otimes SU(2)_L \otimes U(1)_Y$-singlet scalar:
$\phi_{Z} =(1,1,1,0)$. In \cite{hung2}, this complex singlet scalar field
is written as

\be
\label{phiz}
\phi_{Z} = (v_{Z} + \sigma_Z)\,\exp(ia_Z/v_{Z}) \,,
\ee
where $\langle \sigma_Z \rangle =0$ and
$\langle a_Z \rangle =0$ with $\langle \phi_{Z} \rangle = v_{Z}$. The ``angular''
part of this field, $a_Z$, which is an axion-like particle, plays the role
of the acceleron in the model of dark energy of \cite{hung2}. The ``radial'' part
of the field, $\sigma_Z$, could play the part of the inflaton in a
``low-scale'' (i.e. a scale which is less than a typical GUT value) inflationary
scenario \cite{barcelona}.

As shown in \cite{hung2}, the model exhibits a global
$U(1)_{A}^{(Z)}$ symmetry which is spontaneously broken by $\langle
\phi_{Z} \rangle = v_{Z}$ and explicitly broken by the $SU(2)_Z$
instanton effects (making $a_Z$ a Pseudo-Nambu-Goldstone (PNG) boson
instead of a NG boson). The fermions $\psi_{i}^{(Z)}$ acquire masses
through the couplings with $\phi_{Z}$ \cite{hung2}.

The evolution of the $SU(2)_Z$ gauge coupling, $g_Z$, depends on the particle
content listed in (I) and (II). As shown in \cite{hung2}, starting
with an initial value at a scale $\sim 10^{16}\,GeV$ within a factor
of two or three of the corresponding SM values for the gauge couplings,
$\alpha_Z = g_Z^2/4\pi$, remains relatively flat down to O($100\,GeV$),
when $\psi_{i}^{(Z)}$ decouple, and starts to rise until $\alpha_Z =1$
at $\sim 3 \times 10^{-3}\,eV$. This is the scale when $SU(2)_Z$ grows
strong and where the instanton-induced $a_Z$-potential used in the dark
energy model is generated.

\subsection{Representations of $E_6$}

How do the above particle content along with the SM particles fit into
representations of a possible GUT group $E_6$? We first notice that
the fermions $\psi_{i}^{(Z)}$ and the messenger scalar fields
${\tilde{\bm{\varphi}}}_{i}^{(Z)}$ transform as adjoints of
$SU(2)_Z$. As it will become clear below, $\psi_{i}^{(Z)}$  would
fit into adjoint representations, {\bf 78}, of $E_6$. This is
in contrast with SM particles which, as we shall see, are
grouped in fundamental representations {\bf 27}. In order to
see what one really needs, we first observe that the maximal
subgroup of interest in this paper is the following:

\be
\label{subgroup}
E_6 \supset SU(2)_Z \otimes SU(6) \,.
\ee
Notice that the most frequent embedding that one encounters in
GUT models is $E_6 \supset SO(10) \otimes U(1)$, where $SO(10)$
represents the popular GUT model \cite{E6}. Our symmetry breaking path is
different here. Since the dark energy model involves an {\em unbroken}
$SU(2)_Z$, it is the above \ref{subgroup} maximal subgroup
of $E_6$ that we will be concerned with, namely $SU(2)_Z \otimes SU(6)$.
As we shall see below, the subgroup $SU(6)$ contains the SM and is broken
in several steps. At this point,
it is useful to list the maximal subgroups of $SU(6)$, namely
$SU(6) \supset SU(5) \otimes U(1)$,
$SU(6) \supset SU(2) \otimes SU(4) \otimes U(1)$,
and $SU(6) \supset SU(3) \otimes SU(3) \otimes U(1)$.
It is this last embedding which forms the core of our model.
In the next section, we will discuss the symmetry breaking of $E_6$ and
of its subgroups.
We will focus in this paper on a particular
model
\be
\label{SSBchain}
E_6 \stackrel{\tiny{\textsl{M}_{GUT}}}{\longrightarrow} G_1
\stackrel{\tiny{\textsl{M}_{6}}}{\longrightarrow}G_2
\stackrel{\tiny{\textsl{M}_{3L}}}{\longrightarrow}G_3
\stackrel{\tiny{\textsl{M}_{W}}}{\longrightarrow} G_{4}
\ee
where,
\bes
\label{groups}
\be
G_1=SU(2)_Z \otimes SU(6)\,,
\ee
\be
G_2=SU(2)_Z \otimes SU(3)_c \otimes SU(3)_L \otimes U(1)_6\,,
\ee
\be
G_{3}=SU(2)_Z \otimes SU(3)_c \otimes SU(2)_L \otimes U(1)_{Y}\,.
\ee
\be
G_{4} = SU(2)_Z \otimes
SU(3)_c \otimes U(1)_{em}\,,
\ee
\ees
For the moment, we first focus on the
matter representations under $E_6$ and its subgroups.

Let us first look at the decomposition of the two $E_6$
representations which are of interest to us: {\bf 27} and {\bf 78},
under $SU(2)_Z \otimes SU(6)$. What follows is a general
discussion which, for the moment, ignores chirality issues to be
dealt with subsequently. One has
\bes
\be
\label{27}
\textbf{27} = (\textbf{2},\bar{\textbf{6}})+ (\textbf{1},\textbf{15}) \,,
\ee
\be
\label{78}
\textbf{78} = (\textbf{3},\textbf{1})+(\textbf{1},\textbf{35}) +
(\textbf{2},\textbf{20}) \,.
\ee
\ees

A preliminary look at (\ref{78}) reveals the fact that the $SU(2)_Z$
fermions $\psi_{i}^{(Z)}$ should fit into the {\bf 78}
representations, being a triplet under $SU(2)_Z$ and a singlet under
$SU(6)$ (which contains the SM). The next question concerns the SM
fermions and how they fit. As stressed above, one feature of the
dark energy model \cite{hung2} is the fact that SM particles are
{\em singlets} under $SU(2)_Z$. From (\ref{27}), one has
$(\textbf{1},\textbf{15})$ and from (\ref{78}), one has
$(\textbf{1},\textbf{35})$. Since $SU(6)$ will be subsequently
broken down to the SM (in several steps) and since SM particles are
in fundamental representations of $SU(3)_c \otimes SU(2)_L \otimes
U(1)_Y$, it can be seen that $(\textbf{1},\textbf{35})$ (which
belongs to {\bf 78}) cannot contain the SM particles. (In
particular, under $SU(3)_c \otimes SU(3)_L \otimes U(1)$,
$\textbf{35}= (\textbf{1}, \textbf{1})(0)+(\textbf{8},
\textbf{1})(0)+ (\textbf{1}, \textbf{8})(0)+ (\textbf{3},
\bar{\textbf{3}})(-1/\sqrt{3}) +
(\bar{\textbf{3}},\textbf{3})(1/\sqrt{3})$, which does not contain
SM leptons among others.)

We then conclude that $\psi_{i}^{(Z)}$ and
SM fermions fit into {\bf 78} and {\bf 27} representations respectively.
We next present in detail this embedding.

\subsection{Fermion embedding into $E_6$ representations and those
of its subgroups}

The fermions of the model of \cite{hung2} can now be classified under $E_6$ and its
subgroup $SU(2)_Z \otimes SU(6)$ as follows.

\bi

\item {\bf $\psi_{i}^{(Z)}$}:

Since $SU(2)_Z$ is an unbroken, vector-like gauge group (just like QCD), both
left- and right-handed $\psi_{i}^{(Z)}$ transform in the same way (triplets)
under $SU(2)_Z$ as listed in (\ref{psiZ}). In consequence, $\psi_{i,(L,R)}^{(Z)}$
is part of ${\bf 78}^{i}_{L,R}$ which is
\be
\label{78LR}
\textbf{78}^{i}_{L,R} = (\textbf{3},\textbf{1})^{i}_{L,R}+
(\textbf{1},\textbf{35})^{i}_{L,R} + (\textbf{2},\textbf{20})^{i}_{L,R} \,,
\ee
with $i=1,2$. From this, one can readily make the identification
$\psi_{i,(L,R)}^{(Z)} \equiv (\textbf{3},\textbf{1})^{i}_{L,R}$. We will discuss
the masses of these fermions below, in particular we will show how these
fermions can avoid bare masses (being vector-like).

\item {\bf SM fermions}:

SM fermions transform under $SU(2)_Z \otimes SU(3)_c \otimes SU(2)_L \otimes U(1)_Y$
as $q_L=(1,3,2,1/3)$, $u_R=(1,3,1,2/3)$,
$d_R=(1,3,1,-1/3)$, and $l_L=(1,1,2,-1/2)$, $e_R=(1,1,1,-1)$, with the
last entries denoting $Y/2$, the $U(1)_Y$ quantum numbers. Can they fit
into a single {\bf 27} representation of $E_6$?

First, we notice that the parts $(\textbf{2},\bar{\textbf{6}})$ of
{\bf 27} under $SU(2)_Z \otimes SU(6)$ cannot contain SM particles.
(They will acquire a large mass as shown below.) This leaves us with
$(\textbf{1},\textbf{15})$. As stated above the path that we chose
for the breaking of $SU(6)$ is $SU(6) \rightarrow SU(3) \otimes
SU(3) \otimes U(1)$, with $\textbf{15}$ transforming under
$SU(3)\otimes SU(3)$ as $\textbf{15} =
(\bar{\textbf{3}},\textbf{1})+ (\textbf{1}, \bar{\textbf{3}})+
(\textbf{3},\textbf{3})$. To see explicitly how this might represent
the usual SM particles, we rewrite $SU(3)\otimes SU(3)$ as $SU(3)_c
\otimes SU(3)_L$. Let us first start with $\textbf{15}_L$ (which is
part of $\textbf{27}_L$). We have \be \label{15L} \textbf{15}_L =
(\bar{\textbf{3}},\textbf{1})_L+ (\textbf{1}, \bar{\textbf{3}})_L +
(\textbf{3},\textbf{3})_L \,. \ee Using the well-known Weyl
two-component spinors, one has $\psi^{c}_{L} = \sigma_{2}
\psi^{*}_{R}$. This means that $\bar{\textbf{3}}_L$ comes from
$\textbf{3}_R$. Hence (\ref{15L}) contains a colored left-handed triplet
and a color-singlet {\em right-handed} triplet under $SU(3)_L$. In the
next section where we will discuss the symmetry breaking of the
model, it will be seen that $SU(3)_L \supset SU(2)_L$. Therefore
under $SU(2)_L$, (\ref{15L}) contains a left-handed ``quark''
doublet and a right-handed ``lepton'' doublet.

The above discussion reveals two important points. The first is that
(\ref{15L}) (or $\textbf{27}_L$) cannot alone accommodate SM
leptons. The second point is that it contains ``mirror fermions'' in
the form of the right-handed ``lepton'' doublet. We are now forced
to introduce another chiral representation, $\textbf{27}_R$, in
order to have all SM particles per family. One has \be \label{15R}
\textbf{15}_R = (\bar{\textbf{3}},\textbf{1})_R+ (\textbf{1},
\bar{\textbf{3}})_R + (\textbf{3},\textbf{3})_R \,. \ee Let us now
notice that $\bar{\textbf{3}}_R$ transforms as $\textbf{3}_L$. In
consequence, (\ref{15R}) contains the usual left-handed lepton
doublet in addition to an extra right-handed ``mirror'' quark
doublet.

As for $SU(3)_L$ singlets, (\ref{15L}) contains a right-handed singlet ``quark''
while (\ref{15R}) contains a left-handed singlet ``mirror'' quark.

Notice also that $\textbf{27}_L$ and $\textbf{27}_R$ have in addition
to the SM particles and their mirror counterparts, the following
vector-like fermions $(\textbf{2},\bar{\textbf{6}})_{L,R}$. These fermions
can acquire a large mass as we will show below.

In summary, in order to fit the SM fermion spectrum in representations
of $E_6$ which is spontaneously broken according to the chain (\ref{SSBchain}),
we are required to have {\em both} $\textbf{27}_L$ and $\textbf{27}_R$. The
price that on has to pay in this scenario is the existence of mirror
fermions which can be sufficiently heavy in order to escape detection at the
present time. (The phenomenology of these fermions will be presented at the
end of the manuscript.) There is one remaining remark which is worth
mentioning: $E_6$ in our case is a {\em vector-like} model. As such, it
is rather different from the usual chiral versions of $E_6$. However,
weak parity violation involving SM particles is reflected in this case
in the way they are embedded in the above representations.

\ei

\subsection{Embedding of the messenger scalar fields
${\tilde{\bm{\varphi}}}_{i}^{(Z)}$ in $E_6$ representations}

\bi

\item ${\tilde{\bm{\varphi}}}_{i}^{(Z)}$:

The messenger scalar fields ${\tilde{\bm{\varphi}}}_{i}^{(Z)}$ of \cite{hung2}
transform as $(\textbf{3},\textbf{2})$ under $SU(2)_Z \otimes SU(2)_L$
(see (\ref{messenger}). (It is a color-singlet.)
The representation with the lowest dimension that
contains a triplet of $SU(2)_Z$ and a doublet of $SU(2)_L \subset SU(6)$
is $\overline{\textbf{351}}$ which has the following decomposition under
$SU(2)_Z \otimes SU(6)$:
\be
\label{351}
\overline{\textbf{351}}= (\textbf{2},\textbf{6})+ (\textbf{1},\overline{\textbf{21}})
+(\textbf{3},\overline{\textbf{15}})+(\textbf{1},\overline{\textbf{105}})+
(\textbf{2},\textbf{84})\,.
\ee
One can readily see that ${\tilde{\bm{\varphi}}}_{i}^{(Z)}
\in (\textbf{3},\overline{\textbf{15}})$ since
$\overline{\textbf{15}} \supset \textbf{2}$.

\item $\phi_{Z}$:

$\phi_{Z}$ as presented in \cite{hung2} is a singlet under both $SU(2)_Z$
and the SM. The most economical way is for $\phi_{Z}$ to be a {\em singlet}
of $E_6$ as well.

\ei

Next in our discussion is the subject of symmetry breaking which also
includes some issues of fermion masses.

\section{Pattern of symmetry breaking of $E_6$, fermion masses, proton decay}
\label{SSBsect}

\subsection{Breaking of $E_6$}

The pattern of symmetry breaking of $E_6$ that we discuss in this section
is given in (\ref{SSBchain}). We now discuss each step of the breaking chain.
In this discussion, we will keep in mind that $SU(2)_Z$ remains {\em unbroken}.

\bi

\item $E_6 \stackrel{\tiny{\textsl{M}_{GUT}}}{\longrightarrow}
SU(2)_Z \otimes SU(6)$:

To achieve the above breaking, one should find a Higgs representation which
contains a {\em singlet} under $SU(2)_Z \otimes SU(6)$. It is
\ba
\label{650}
\textbf{650}&=&(\textbf{1},\textbf{1})+(\textbf{1},\textbf{35})+(\textbf{2},
\textbf{20})+(\textbf{3},\textbf{35})+(\textbf{2},\textbf{70}) \nonumber \\
&& +(\textbf{2},\overline{\textbf{70}})+(\textbf{1},\textbf{189}) \,.
\ea
From (\ref{650}), it follows that
\be
\label{650vev}
\langle \textbf{650} \rangle = \langle (\textbf{1},\textbf{1}) \rangle \neq 0 \,
\ee
achieves the desired breaking
$E_6 \stackrel{\tiny{\textsl{M}_{GUT}}}{\longrightarrow}
SU(2)_Z \otimes SU(6)$. In this step, the
$E_6/SU(2)_Z \otimes SU(6)$ gauge bosons acquire a mass of order $M_{GUT}$.

\item $SU(2)_Z \otimes SU(6) \stackrel{\tiny{\textsl{M}_{6}}}{\longrightarrow}
SU(2)_Z \otimes SU(3)_c \otimes SU(3)_L \otimes U(1)_6$:

Notice that, in what follows, our normalization for the
$U(1)_6$ quantum numbers which will be used in Section (\ref{RG})
differs from the ones used in \cite{Slansky} by a factor
of $1/(2\sqrt{3})$.
The breaking of $SU(6)$ down to $SU(3)_c \otimes SU(3)_L
\otimes U(1)_6$ is achieved by looking at the decomposition:
\ba
\label{35}
\textbf{35}& =& (\textbf{1},\textbf{1})(0)+(\textbf{8},\textbf{1})(0)
+(\textbf{1},\textbf{8})(0) \nonumber \\
&&+ (\textbf{3},\bar{\textbf{3}})(-1/\sqrt{3})+
(\bar{\textbf{3}},\textbf{3})(1/\sqrt{3}) \,.
\ea
So
\be
\label{35vev}
\langle \textbf{35} \rangle = \langle (\textbf{1},\textbf{1})(0) \rangle
\neq 0 \,,
\ee
will give $SU(6) \stackrel{\tiny{\textsl{M}_{6}}}{\longrightarrow}
SU(3)_c \otimes SU(3)_L \otimes U(1)_6$. Since $(\textbf{1},\textbf{35})$
is also contained in {\bf 650}, one might use the same field to achieve
both breaking with $\langle (\textbf{1},\textbf{35}) \rangle \sim M_6 <
\langle (\textbf{1},\textbf{1}) \rangle \sim M_{GUT}$. We now have
\be
\label{650vev2}
\langle \textbf{650} \rangle = \langle (\textbf{1},\textbf{1}) \rangle
(\sim M_{GUT}) +  \langle (\textbf{1},\textbf{35}) \rangle (\sim M_6)\,.
\ee
The $SU(6)/SU(3)_c \otimes SU(3)_L \otimes U(1)_6$ gauge bosons acquire
a mass of order $M_6$.

Since SM fermions and their mirror counterparts belong to a {\bf 15} of
$SU(6)$, the $U(1)_6$ generator, which is a diagonal generator of
$SU(6)$, $T_{35}$, has the normalization $Tr T_{35}^2 =2$ (with
the usual convention that, for the fundamental representation, one
has $Tr T_{i}^2 =1/2$). We will see below that the appropriate
coefficient which multiplies $T_{35}$ is $C_6 = 2/\sqrt{3}$ in
order to make connection with the SM $U(1)_Y$ quantum numbers.

\item $SU(3)_L \otimes U(1)_6 \stackrel{\tiny{\textsl{M}_{3L}}}{\longrightarrow}
SU(2)_L \otimes U(1)_{Y}$:

In this breaking, the generator of $U(1)_Y$ is a linear combination
of $T_{35}$ and a diagonal generator of $SU(3)_L$, namely $T_{8L}$, which
is actually one of the diagonal generators of the original $SU(6)$. It is
\be
\label{Y}
Y/2 = C_6 T_{35} + C_{3L} T_{8L} = (2/\sqrt{3})T_{35} +
(1/\sqrt{3})T_{8L} \,,
\ee
in order to get the correct weak hypercharge
quantum numbers for the SM particles as we shall see below. The Higgs
representation that can accomplish the above breaking should be
a singlet under $SU(2)_L$ and should have $Y/2 =0$.

The Higgs representation which satisfies the above criterion is the
following \be \label{21} \textbf{21}
=(\textbf{6},\textbf{1})(-1/\sqrt{3})+
(\textbf{1},\textbf{6})(1/\sqrt{3})+ (\textbf{3},\textbf{3})(0) \,.
\ee

The component $(\textbf{1},\textbf{6})(1/\sqrt{3})$ is the one that
satisfies the desired requirement. Since {\bf 6} is a symmetric
second-rank tensor of $SU(3)_L$, it is straightforward (see e.g.
\cite{LiFong}) to obtain the desired breaking $SU(3)_L \otimes
U(1)_6 \stackrel{\tiny{\textsl{M}_{3L}}} {\longrightarrow}SU(2)_L
\otimes U(1)_{Y}$ when $\langle (\textbf{1},\textbf{6})(1/\sqrt{3})
\rangle \neq 0$. One can also see that the $U(1)_Y$ gauge boson
$B^{\mu}$ is found to be (see e.g. \cite{PUT1} and \cite{PUT2}) \be \label{B}
B^{\mu} = \cos \theta_L A_{35}^{\mu} + \sin \theta_L A_{8L}^{\mu}
\,, \ee where \be \label{thetal} \cos \theta_L =
\frac{g_{3L}\,C_6}{\sqrt{g_{3L}^2\,C_6^2 + g_{6}^2\,C_{3L}^2}}\,;\,
\sin \theta_L = \frac{g_{6}\,C_{3L}} {\sqrt{g_{3L}^2\,C_6^2 +
g_{6}^2\,C_{3L}^2}}\,. \ee

The following gauge bosons obtain masses of O($M_{3}$):
an $SU(2)_L$ doublet (which absorbs the doublet Nambu-Goldstone (NG) bosons
of the {\bf 6} scalar) and the combination which is orthogonal to
$B^{\mu}$ (which absorbs the imaginary part of $SU(2)_L$ singlet
scalar).

\item $SU(2)_L \otimes U(1)_{Y} \stackrel{\tiny{\textsl{M}_{W}}}{\longrightarrow}
U(1)_{em}$

This last step is accomplished in the usual manner, namely by the use of
a $SU(2)_L$ complex Higgs doublet belonging to
$(\textbf{1},\bar{\textbf{3}}) \subset (\textbf{1},\textbf{15}) \subset
\textbf{27}$.

\ei

\subsection{Fermion masses}
\label{fmass}

The main focus in this section is to discuss which fermion might be ``heavy''
and which might be ``light''. The main purpose is to know when to include
or exclude a certain fermion from the RG evolution of the gauge couplings.

\bi

\item {\bf 27}:

Let us remember that a mass term can be written in terms of two-component
Weyl spinors as $\psi^{T}_{1,(L,R)}\,\sigma_{2}\,\psi_{2,(L,R)}$. Also, one
can solely use left-handed Weyl spinors by recalling that
$\psi^{c}_{L} = \sigma_{2} \psi^{*}_{R}$.
Since our model contains $\textbf{27}_{L,R}$ with
now $\textbf{27}_{L}^{c} = \sigma_{2}\, \textbf{27}_{R}^{*}$, we
obtain the following Lorentz-invariant combinations
\bes
\be
\label{271}
\textbf{27}_{L}^{T} \sigma_{2} \textbf{27}_{L} \sim
\textbf{27} + \textbf{351} +\textbf{351}^{'} \,,
\ee
\be
\label{272}
\textbf{27}_{L}^{c,T} \sigma_{2} \textbf{27}_{L}^{c} \sim
\textbf{27} + \textbf{351} +\textbf{351}^{'} \,,
\ee
\be
\label{273}
\textbf{27}_{L}^{c,T}\sigma_{2} \textbf{27}_{L} \sim \textbf{1}
+ \textbf{78} + \textbf{650} \,,
\ee
\ees
where the right-hand sides of Eqs. (\ref{271}, \ref{272}, \ref{273})
denote the resulting representations under $E_6$.

Since (\ref{273})
contains a singlet, one can avoid a gauge-invariant bare mass term
by having e.g. a discrete symmetry such that one can assign
$\textbf{27}_{L} \rightarrow \textbf{27}_{L}$ and
$\textbf{27}_{R} \rightarrow -\textbf{27}_{R}$. In consequence,
(\ref{271}) and (\ref{272}) are even while (\ref{273}) is odd under
that symmetry. This prevents {\bf 27} from having a bare mass term.
As a result, the possible Higgs representations that appear on the
right-hand side of (\ref{271}) and (\ref{272}) have even ``parity''
while those on the right-hand side of (\ref{273}) should possess
odd ``parity''.

The next step is to examine which Higgs representation will be
appropriate to use to give masses to the fermions which belong to
$\textbf{27}_{L,R}$. We must first understand which fermion bilinears
(relevant for the mass terms) are contained in
(\ref{271},\ref{272},\ref{273}). In order to do so, we must first recall
the particle contents of $\textbf{27}_{L,R}$. In particular,
we would like to know the $SU(2)_L \otimes U(1)_Y$ quantum numbers
of the SM fermions and their mirror counterparts contained in
$\textbf{27}_{L,R}$, namely $(\textbf{1},\textbf{15})_{L,R}
\in \textbf{27}_{L,R}$.

From hereon, we will list the particle contents of $\textbf{15}_{L,R}$
under $SU(2)_L \otimes U(1)_Y$, making use of
$\psi^{c}_{L,R} = \sigma_{2} \psi^{*}_{R,L}$. We have
\ba
\label{15L2}
\textbf{15}_L &=& (\bar{\textbf{3}},\textbf{1})_L+ (\textbf{1},
\bar{\textbf{3}})_L + (\textbf{3},\textbf{3})_L \nonumber \\
&=& (u^{c}_{L}, l^{M,c}_L + e^{c}_L, q_L + d^{M}_L) \,, \ea where
the particle identifications follow from the quantum numbers coming
from the pattern of symmetry breaking of the last section and where
$u^{c}_{L}$ represents an $SU(2)_L$ singlet right-handed up-quark,
$l^{M,c}_L$ a mirror right-handed lepton doublet, $e^{c}_L$ a
$SU(2)_L$ singlet right-handed charged lepton, $ q_L$ a left-handed
quark doublet, and $d^{M}_L$ a mirror $SU(2)_L$ singlet right-handed
down-quark. Similarly, one can carry out the same exercise with
(\ref{15R}) to get \be \label{15R2} \textbf{15}^{c}_L = (u^{M}_{L},
l_L + e^{M}_L, q^{M,c}_L + d^{c}_L) \,. \ee From
(\ref{15L2},\ref{15R2}), it is straightforward to see that the
following fermion bilinears which are relevant for the mass terms.
\bi

\item SM fermion bilinears:

1) Up-quark mass term $u^{c,T}_{L}\,\sigma_2 \,q_L$ from
$\textbf{15}_L^{T} \,\sigma_2 \, \textbf{15}_L$.

2) Down-quark mass term $d^{c,T}_{L}\,\sigma_2 \,q_L$, charged lepton mass
term $e^{c,T}_L\,\sigma_2 \,l_L$ from
$\textbf{15}^{c,T}_L \,\sigma_2 \, \textbf{15}_L$
and $\textbf{15}^{T}_L \,\sigma_2 \, \textbf{15}^{c}_L$.

\item Mirror fermion bilinears:

1) Mirror Up-quark mass term $u^{M,T}_{L}\,\sigma_2 \,q^{M,c}_L$ from
$\textbf{15}^{c,T}_L \,\sigma_2 \, \textbf{15}^{c}_L$.

2) Mirror Down-quark mass term $d^{M,T}_{L}\,\sigma_2 \,q^{M,c}_L$,
charged lepton mass term $e^{M,T}_L\,\sigma_2 \,l^{M,c}_L$ from
$\textbf{15}^{c,T}_L \,\sigma_2 \, \textbf{15}_L$
and $\textbf{15}^{T}_L \,\sigma_2 \, \textbf{15}^{c}_L$.

\ei

Notice that we have not touched the issue of neutrino masses in the above discussion.
This will be dealt with at the end of this section. We first examine the Higgs
scalars which can couple to the above fermion bilinears.

\bi

\item The Higgs representation that can give masses to the (SM and mirror)
Up-quarks can be inferred by looking at the product
$(\bar{\textbf{3}},\textbf{1})(-1/\sqrt{3}) \times (\textbf{3},\textbf{3})(0)
\supset (\textbf{1},\textbf{3})(-1/\sqrt{3})$, where the last entries refer to
the $U(1)_6$ quantum numbers. In consequence, one needs a Higgs scalar
which transforms as $(\textbf{1},\bar{\textbf{3}})(1/\sqrt{3}) \subset
\textbf{15} \subset \textbf{27}$. We shall call this field
$\phi(\textbf{27})$ whose $SU(2)_L$ doublet which belongs to
$(\textbf{1},\bar{\textbf{3}})(1/\sqrt{3})$ is the one that develops a
non-vanishing V.E.V. resulting in the breakdown of
$SU(2)_L \otimes U(1)_Y$. We shall assume that the mirror fermions
all have masses of $O(>200\,GeV)$. Notice that $\phi(\textbf{27})$
has even ``parity'' as discussed above.

\item For the (SM and mirror) charged leptons and Down-quarks, one looks at the
product $(\textbf{1},\bar{\textbf{3}})(1/\sqrt{3}) \times
(\textbf{1},\textbf{3})(-1/\sqrt{3}) \supset (\textbf{1},\textbf{8})(0)$
and $(\textbf{3},\textbf{3})(0) \times
\bar{(\textbf{3}},\bar{\textbf{3}})(0) \supset (\textbf{1},\textbf{8})(0)$.
This comes from $\textbf{15}^{c,T}_L \,\sigma_2 \, \textbf{15}_L$ and
contains terms such as $e^{c,T}_L\,\sigma_2 \,l_L$, etc...The obvious
choice of the Higgs representation which can couple to these
bilinears is  $(\textbf{1},\textbf{8})(0) \subset (\textbf{1},\textbf{35})
\subset \textbf{78}$. This Higgs field is denoted by  $\phi(\textbf{78})$
The extra $SU(2)_L$ Higgs doublet is contained in
$\textbf{8} = \textbf{1} + \textbf{2} + \bar{\textbf{2}} + \textbf{3}$.
Again we shall assume that the mirror fermions
all have masses of $O(>200\,GeV)$. According to the discussion above,
$\phi(\textbf{78})$ has odd ``parity''.
\ei

The above discussions point to the fact that we need at least two
Higgs doublets in our model: one for the Up sector and one for the Down sector.

The next question concerns the mass of $(\textbf{2},\bar{\textbf{6}})_{L,R} \in
\textbf{27}_{L,R}$. Notice that
$(\textbf{2},\bar{\textbf{6}})^{c}_{L} \times (\textbf{2},\bar{\textbf{6}})_{L}
\supset (\textbf{1}, \textbf{1})$. Looking at (\ref{273}, \ref{650}), one
notices that $(\textbf{1}, \textbf{1}) \subset \textbf{650}$ and it is
the V.E.V. of this component that breaks $E_6$ down to $SU(2)_Z \otimes SU(6)$
at $M_{GUT}$. Therefore, $(\textbf{2},\bar{\textbf{6}})_{L,R}$ can obtain a
mass of $O(M_{GUT})$ by coupling to the {\bf 650} Higgs field which has
odd ``parity''.

Finally, we now come to the topic of neutrino masses. First, we
notice that the mirror lepton doublet contains a right-handed
neutral lepton. Could it be the right-handed neutrino? Normally, the
right-handed neutrino is viewed as a singlet of $SU(2)_L \otimes
U(1)_Y$. For example, in left-right symmetric models \cite{LR}, the
right-handed neutrino is a member of a right-handed doublet of the
gauge group $SU(2)_R$ and therefore is naturally a $SU(2)_L \otimes
U(1)_Y$ singlet. In our case, the right-handed neutral lepton is a
{\em member} of an $SU(2)_L$ doublet but its partner is however {\em
not} the SM charged lepton but its mirror counterpart. Since it does
not interact at tree level with a SM charged lepton, it could play
the role of the right-handed neutrino. If this is the case, it
follows that the Dirac neutrino mass term cannot come from a coupling to
the SM Higgs doublet(s), the reason being that $(\bar{\nu}_R,
\bar{e}^{M}_R) \times (\nu_L, e_L)$ is an $SU(2)_L$ singlet. From
(\ref{273}), we notice that the previous term comes from
$\textbf{27}_{L}^{c,T}\sigma_{2} \textbf{27}_{L}$ which contains an
$E_6$ singlet. A {\em Dirac} neutrino mass term can then arises from
$\textbf{27}_{L}^{c,T}\sigma_{2} \textbf{27}_{L}\,\phi_S
(\textbf{1})$, where $\phi_S(\textbf{1})$ is an $E_6$ singlet and
has odd ``parity''. The Dirac neutrino mass is then found to be
$m_{\nu} = g_{\nu} \langle \phi_S(\textbf{1}) \rangle$. The
interesting fact is that Dirac neutrino masses can be small not by
fine-tuning the Yukawa coupling in the case where one simply adds a
right-handed neutrino to the SM, but by having a small V.E.V.
$\langle \phi_S(\textbf{1}) \rangle$. The implication of this
scenario in terms of the
see-saw mechanism will be dealt with elsewhere \cite{hung4}. 
Notice that the
aforementioned coupling also gives a mixing between the mirror
quarks and their SM counterparts, and similarly between the charged
mirror and SM leptons. However, this mixing is highly suppressed
with an angle being proportional to $m_{\nu}/(m_q-m_{q^M}, m_l-m_{l^M})$.

\item {\bf 78}:

The fermions $\textbf{78}^{i}_{L,R}$ decompose under $SU(2)_Z
\otimes SU(6)$ as shown in (\ref{78LR}). Let us recall that
$\psi_{i}^{(Z)}= (\textbf{3}, \textbf{1})^{i} \in \textbf{78}^{i}$
where $i=1,2$. Let us also recall that the model proposed in
\cite{hung2} contains a global $U(1)_{A}$ symmetry under which one
has the following transformations: $\psi^{(Z)}_{L,i} \rightarrow
e^{-i\alpha}\,\psi^{(Z)}_{L,i}$, $\psi^{(Z)}_{R,i} \rightarrow
e^{i\alpha}\,\psi^{(Z)}_{R,i}$, which also apply to
$\textbf{78}^{i}_{L,R}$. A Dirac mass term of the form
$\textbf{78}^{c,T}_{i,L} \sigma_2 \textbf{78}_{i,L}$ would carry a
phase $e^{-2\,i\alpha}$ under that transformation. It follows that
the Higgs field which couples to the aforementioned fermion bilinear
should have a phase $e^{+2\,i\alpha}$ under the same transformation.
This cannot be the Higgs fields that couple to the {\bf 27} as we
have discussed above. What could the minimal Higgs additions to the
previous choices be? First notice that \be \label{7878}
\textbf{78}^{c,T}_{i,L} \sigma_2 \textbf{78}_{i,L} \sim \textbf{1} +
\textbf{78} + \textbf{650} \,, \ee and \bes \label{78prod} \be
(\textbf{2},\textbf{20}) \times (\textbf{2},\textbf{20}) \supset
(\textbf{1},\textbf{1}) + (\textbf{1},\textbf{35}) \,, \ee \be
(\textbf{1},\textbf{35}) \times (\textbf{1},\textbf{35}) \supset
(\textbf{1},\textbf{1}) + (\textbf{1},\textbf{35}) \,, \ee \be
(\textbf{3},\textbf{1}) \times (\textbf{3},\textbf{1}) =
(\textbf{1},\textbf{1}) + (\textbf{3},\textbf{1}) +
(\textbf{5},\textbf{1})\,. \ee \ees In (\ref{78prod}), the
$(\textbf{1},\textbf{1})$ can be most conveniently the $E_6$ singlet
$\textbf{1}$, and $(\textbf{1},\textbf{35})$ can be part of the
$\textbf{78}$ on the right-hand side of Eq. (\ref{7878}). We shall
denote these scalars by $\phi(\textbf{1})$ and $\phi(\textbf{78})$,
both of which acquire a phase $e^{+2\,i\alpha}$ under the $U(1)_{A}$
transformations.

In this context, we readily identify $\phi(\textbf{1}) \equiv
\phi_Z$ with $\phi_Z$ being the singlet scalar field used in
the model of dark energy and dark matter of
\cite{hung1}, \cite{hung2}. Let us recall that the imaginary part of $\phi_Z$
plays the role of the acceleron \cite{hung2} and the real part plays the role
of the inflaton \cite{barcelona}. Its V.E.V. $\langle \phi_{Z} \rangle = v_{Z}$
gives a common mass of $O(m_{\psi^{(Z)}})$ to
$(\textbf{2},\textbf{20})$, $(\textbf{1},\textbf{35})$,
and $(\textbf{3},\textbf{1}) \equiv \psi^{(Z)}$ through a coupling
$\textbf{78}^{c,T}_{i,L} \sigma_2 \textbf{78}_{i,L}\, \phi_Z$.
Furthermore, a coupling of the form $\textbf{78}^{c,T}_{i,L} \sigma_2
\textbf{78}_{i,L}\,\phi(\textbf{78})$ gives a common mass to
$(\textbf{2},\textbf{20})$ and $(\textbf{1},\textbf{35})$ when
$\langle (\textbf{1},\textbf{35}) \rangle \sim O(M_{6})$.
From \cite{hung2}, one expects $m_{\psi^{(Z)}} \sim O(200\,GeV)$
which is much smaller than $M_6$. In consequence, the fermions
$(\textbf{2},\textbf{20})$ and $(\textbf{1},\textbf{35})$ of $\textbf{78}$
having a mass of order $M_6$,
are {\em much heavier} than $(\textbf{3},\textbf{1}) \equiv \psi^{(Z)}$.

\ei

In summary, we have shown above how the $(\textbf{2},\bar{\textbf{6}})
\in \textbf{27}$, $(\textbf{2},\textbf{20}) \in \textbf{78}$,
and $(\textbf{1},\textbf{35}) \in \textbf{78}$ can become very heavy
while the SM particles and their counterparts as well as
$\psi^{(Z),i}$ can remain ``light''.

\subsection{Proton decay}

The spontaneous breakdown of the subgroup $SU(6)$ will in general
induce proton decays with a lifetime being proportional to
$M_{6}^{4}/\alpha_{6}^2$. But one first has to make sure from
a group-theoretical point of view that operators responsible
for the proton decay exist in the model.

First, let us recall that the V.E.V. of a {\bf 35} Higgs breaks
$SU(6)$ down to $SU(3)_c \otimes SU(3)_L \otimes U(1)_6$ giving
masses of $O(M_6)$ to the $(\textbf{3},\bar{\textbf{3}})(-1/\sqrt{3})$
and $(\bar{\textbf{3}},\textbf{3})(1/\sqrt{3})$
gauge bosons (see Eq.(\ref{35})).
Under $SU(2)_L$, these gauge bosons transform as
$(\textbf{3},\bar{\textbf{3}})(-1/\sqrt{3}) =
(\textbf{3},\bar{\textbf{2}})(-1/\sqrt{3})
+(\textbf{3},\textbf{1})(-1/\sqrt{3})$ and
$(\bar{\textbf{3}},\textbf{3})(1/\sqrt{3})=
(\bar{\textbf{3}},\textbf{2})(1/\sqrt{3})
+ (\bar{\textbf{3}},\textbf{1})(1/\sqrt{3})$.
For definiteness, let us denote these gauge bosons by
$U_{\mu} \equiv (\textbf{3},\bar{\textbf{3}})(-1/\sqrt{3})$
and $\bar{U}_{\mu} \equiv (\bar{\textbf{3}},\textbf{3})(1/\sqrt{3})$.
From Eqs.(\ref{15L}, \ref{15L2}), one can see that the relevant
transitions for proton decay are the following: $q_L \rightarrow
e^{c}_L,\,u^{c}_{L}$. This means that, in terms of representations,
one has $(\textbf{3},\textbf{3})(0)_L \rightarrow
(\textbf{1}, \bar{\textbf{3}})(1/\sqrt{3})_L,\,
(\bar{\textbf{3}},\textbf{1})(-1/\sqrt{3})_L$.

\bi

\item $\psi^{1}_L \equiv (\textbf{3},\textbf{3})(0)_L$,
$\psi^{2}_L \equiv (\textbf{1}, \bar{\textbf{3}})(1/\sqrt{3})_L$:

The current
$\bar{\psi^{2}}_L \gamma_{\mu} \psi^{1}_L \sim (\textbf{3},
\bar{\textbf{3}})(-1/\sqrt{3})$
couples to $\bar{U}^{\mu}$. It is straightforward to see that
$g_{6}\, \bar{\psi^{2}}_L \gamma_{\mu} \psi^{1}_L \, \bar{U}^{\mu}$
contains the coupling of $q_L$ to $e^{c}_L$.

\item $\psi^{1}_L \equiv (\textbf{3},\textbf{3})(0)_L$,
$\psi^{2}_L \equiv (\bar{\textbf{3}},\textbf{1})(-1/\sqrt{3})_L$:

The current
$\bar{\psi^{2}}_L \gamma_{\mu} \psi^{1}_L \sim (\bar{\textbf{3}},
\textbf{3})(1/\sqrt{3})$
couples to $U^{\mu}$. Here $g_{6} \, \bar{\psi^{2}}_L \gamma_{\mu}
\psi^{1}_L \, U^{\mu}$ contains the coupling of $q_L$ to
$u^{c}_{L}$.

\item The above interactions describe for instance a proton decay process
such as $p \rightarrow e^{+} \pi^{0}$ by the exchange of the U-boson with mass
of $O(M_6)$. Some numerical estimates are given in the next section.

\ei

\section{Renormalization group analysis}
\label{RG}

In this section we study the evolution of the coupling constants,
associated to the symmetry breaking scheme introduced in section
\ref{SSBsect}. We split our analysis into three parts. The former
contains the general equations describing the evolution of the
couplings at each step of the symmetry breaking process, along
with the generators of the $U(1)$ groups. A formal expression for
$\sin^2\theta_W$ is also derived. In the second part, we expose
our numerical analysis of the problem, where  threshold effects
are properly taken into account. In the last part
we present our results.

\subsection{General analysis}

In this section, we focus on the gauge group $SU(6)$ and its
breaking down to the standard model subgroups, according to the
scheme \ba \label{SSBSU(6)} && SU(6)
\stackrel{\tiny{\textsl{M}_{6}}}{\longrightarrow}SU(3)_c \otimes
SU(3)_L \otimes U(1)_6  \stackrel
{\tiny{\textsl{M}_{3L}}}{\longrightarrow} \\
&& \stackrel{\tiny{\textsl{M}_{3L}}}{\longrightarrow}SU(3)_c \otimes
SU(2)_L \otimes U(1)_{Y}
\stackrel{\tiny{\textsl{M}_{W}}}{\longrightarrow} SU(3)_c \otimes
U(1)_{em}\,. \nonumber
 \ea
\begin{itemize}
\item At mass scale $M_6$, the couplings satisfy the following
condition \be
\label{gM6}g_{3}^2(M_6)=g_{3L}^2(M_6)=g_{1,6}^2(M_6)=g_6^2(M_6)\,,\ee
where, with an obvious notation, $g_3$, $g_{3L}$ and $g_{1,6}$
denote the couplings corresponding respectively to  $SU(3)_c$,
$SU(3)_{L}$ and $U(1)_6$.

The abelian $U(1)_6$ group is associated to the unbroken diagonal
generator, $T_{35}$, of $SU(6)$, which does not belong to
$SU(3)_c\times SU(3)_{L}$.

 For $M_{3L}\leq E\leq M_6$, the solutions to renormalization
group equations read
 \bes\label{RG6}
\be
\frac{1}{g_{3L}^2(M)}=\frac{1}{g_6^2(M_6)}+ 2\,b_{3L}\ln\frac{M}{M_6}\,
\ee
\be \frac{1}{g_{1,6}^2(M)}=\frac{1}{g_6^2(M_6)}+ 2\,b_{1,6}
\ln\frac{M}{M_6}\, \ee 
\be \frac{1}{g_{3}^2(M)}=\frac{1}{g_6^2(M_6)}+
2\,b_{3}\ln\frac{M}{M_6}\,, \ee \ees where the coefficients $b_i$ are
related to the beta function and their explicit expression will be
given in the next subsection.

\item At mass scale $M_{3L}$, the gauge group $SU(3)_L$ breaks
down to $SU(2)_L$ and recombines with $U(1)_6$, in order to give
the weak hypercharge $U(1)_Y$ group.

 The generator of $U(1)_Y$ is
given by the linear combination \be\label{T_Y}T_Y= C_6\,
T_{35}+C_{3L}\,T_{3L}\,,\ee where
 $T_{35}$ has been introduced before and $T_{3L}$ is the
diagonal generator, $T_8$, of $SU(3)_L$. The coefficients
appearing in the expression assume the explicit values \be
C_{6}=2/\sqrt{3} \hspace{0.6cm} \textrm{and}\hspace{0.6cm}
C_{3L}=1/\sqrt{3}\,.\ee

Therefore, the matching conditions for the couplings at $M_{3L}$
read \bes\label{gM3L} \be g_{2}^2(M_{3L})=g_{3L}^2(M_{3L})\,,\ee
\be \frac{1}{g_Y^2(M_{3L})}=\frac{1/3}{g_{3L}^2(M_{3L})}+
\frac{4/3}{g_{1,6}^2(M_{3L})}\,,\ee
\ees while their evolution,  for $M_Z\leq E\leq M_{3L}$, is
governed by \bes \label{RG3L}
\be
\frac{1}{g_{Y}^2(M)}=\left[\frac{1/3}{g_{3L}^2(M_{3L})}+\frac{4/3}
{g_{1,6}^2(M_{3L})}\right]+ 2\,b_{Y}\ln\frac{M}{M_{3L}}\, \ee
\be \frac{1}{g_{2}^2(M)}=\frac{1}{g_{3L}^2(M_{3L})}+
2\,b_{2}\ln\frac{M}{M_{3L}}\, \ee 
\be \frac{1}{g_{3}^2(M)}=\frac{1}{g_6^2(M_6)}+
2\,b_{3}\ln\frac{M}{M_6}\,.\ee  \ees

\item Finally, at the electro-weak  scale $M_Z$ the standard model
groups break down  to $S(3)_c$ and $U(1)_{em}$, with the well
known relations \be Q=T_Y+T_{2L}\,,\ee ($T_{2L}$ being the
diagonal generator  $T_{3}$ of $SU(2)_L$) for the electromagnetic
charge generator and
\be\label{gMZ}\frac{1}{e^2(M_{Z})}=
\frac{1}{g_2^2(M_{Z})}+\frac{1}{g_{Y}^2(M_{Z})}\,.\ee
\end{itemize}
Combining Eq.s
(\ref{gM6}-\ref{gM3L}-\ref{gMZ}-\ref{RG6}-\ref{RG3L}) and  using
the standard $\overline{\textrm{MS}}$ definition for
$\sin^2\theta_W$, i.e. \be
\sin^2\theta_W(M_Z)=\frac{e^2(M_Z)}{g_{2}^2(M_Z)}\,,\ee we obtain
the following formula
\be \label{sin2}
\sin^2\theta_W(M_Z)= \frac{3}{8}
\small{\left\{1- 8 \pi
\alpha(M_Z)
   \left[K\ln\frac{M_Z}{M_{3L}}+K'
\ln\frac{M_{3L}}{M_6}\right]\right\}}  \ee
where \bes \be \alpha(M_Z)=\frac{e^2(M_Z)}{4\pi}\,,\ee
\be K=b_Y-\frac{5}{3}\,b_2\,,\ee
\be K'= \frac{4}{3}\,(b_{1,6}-b_{3L})\,.\ee\ees The overall factor
$3/8$ represents the value of $\sin^2\theta_W$ at the mass scale
$M_{3L}$, where the weak hypercharge group, $U(1)_Y$, first
appears in the symmetry breaking process. The fact that we obtain
the same value $3/8$ as in the $SU(5)$ GUT theory (though at a
different mass threshold) depends on the particle contents of  the
two models: indeed, the mirror fermions have identical quantum
numbers to the SM fermions. (One example of how unconventionally
charged particles affect $\sin^2\theta_W^0$ can be found in
\cite{PUT2}).

\subsection{Numerical analysis}

The behavior of the couplings is analyzed, as the energy
increases, starting from  $\Lambda_Z\sim 3\times 10^{-3}\,
\textrm{eV}$, till almost the Planck scale $M_{Planck}\sim 1.2
\times 10^{19}\gev$. In the following, we first define the
equations used in our analysis, then we summarize the particle
content and show how it affects the evolution, leaving the results
and their discussion to the next section.

The evolution of all the couplings (except from $g_Z$, which we
are going to discuss separately) is well described by the RG
equation at one-loop \cite{beta}
\begin{equation}\label{1loop}
\frac{d \alpha_i}{dt}=-\,\frac{b_i}{2 \pi}\, \alpha_i^2
\end{equation}
where $t\equiv \ln\mu$ and $\alpha=g^2/4\pi$. For a general
product of gauge groups $G_1\otimes G_2\otimes\ldots$, the
coefficients $b_i$ read
\begin{equation}\label{b}
b_i=\frac{11}{3}\;C_2(G_i)-\frac{2}{3}\;T(F_i)\prod_{j\neq i}
d(F_j)-\frac{1}{3}\;T(S_i)\prod_{j\neq i}d(S_j)\, .
\end{equation}
with $C_2$ the quadratic Casimir operator of the group $G_i$,
acting on the adjoint representation,  and $T$ and $d$ being,
respectively, the Dynkin index and the dimension of the chiral
fermion ($F$) and complex scalar ($S$) representations. Since at
very low energies and till roughly the electro-weak scale, only
$\alpha_Z$ is running, our
 initial inputs 
 are those associated to the Standard Model
 couplings at mass scale $M_Z$,
   given by the
experimental data
 \bes \label{inputs}
\be \alpha_Y(M_Z^2)=\frac{\alpha(M_Z^2)}{\cos^2\theta_W(M_Z^2)}\,,\ee
\be \alpha_2(M_Z^2)=
\frac{\alpha(M_Z^2)}{\sin^2\theta_W(M_Z^2)} \,,\ee
\be \alpha_3(M_Z^2)\,, \ee\ees with the $\overline{\textrm{MS}}$ values
\cite{Pdg}
\ba\label{exp}&&1/\alpha(M_Z^2)=127.906(19)\,,\hspace{0.6cm}
\alpha_3(M_Z^2)=0.1213(18)\nonumber
\\ &&\sin^2\theta_W(M_Z^2)|_{exp}=0.23120(15)\nonumber \, .\ea

We now  consider $\alpha_Z=g_Z^2/4\pi$, where $g_Z$
is the $SU(2)_Z$ gauge coupling. Its evolution
 has been already studied in
  \cite{hung2}, for a  range of energies, roughly covering
$\Lambda_Z\leq E\leq M_6$. In this interval, which extends to very
low values, the two-loop approximation turns out  to be  more
accurate. Therefore, \be\label{2loop}
\frac{d\alpha_{Z}}{dt}=-\frac{b^{0}_Z}{2\pi}\,\alpha_Z^{2}-\frac{b^{1}_Z}{8\pi^2}\,\alpha_{Z}^3\,,\ee
where general expressions for the coefficients $b^{0}_Z$ and
$b^{1}_Z$ can be found in \cite{beta} and the initial value is
assumed to be $\alpha_Z(\Lambda_Z)\sim 1$. As concerns higher
energies, above $M_6$,  the one-loop approximation is sufficiently
reliable and we will use Eq. (\ref{1loop}-\ref{b}), taking,  as an
input, the value $\alpha_Z(M_6)$ resulting from Eq. (\ref{2loop}).

The next step consists in calculating explicitly the coefficients
$b_i$. In order to accomplish this, we need to know the
transformation properties, under the gauge groups, of all the
particles involved  at each step of the symmetry breaking process.
First of all, we list all the $E_6$ representations which enter
our analysis:
\begin{itemize}
\item fermions: three $\textbf{27}_{L,R}$ and two $\textbf{78}_{L,R}$;
\item scalars: \begin{itemize}
\item Higgs fields: one $\textbf{650}$, two $\textbf{78}$, one
$\textbf{351}$ and one $\textbf{27}$;
\item messenger fields: two $\overline{\textbf{351}}$.
\end{itemize}
\end{itemize}
Next, we identify  four regions, which are characterized by
different symmetry groups and whose boundaries are determined by
Eq. (\ref{SSBSU(6)}) :
\begin{enumerate}
\item $\Lambda_Z \leq E \leq M_Z\,$,
\item $M_Z \leq E \leq M_{3L}\,$,
\item $M_{3L} \leq E \leq M_{6}\,$,
\item $M_{6} \leq E \leq M_{GUT}\,$.
\end{enumerate}
Within each of them, we will analyze in detail the threshold
effects,  due to the presence of particles with different masses.
(The behavior of $\alpha_Z$ in the interval of energies between
$\Lambda_Z$ and $M_6\,$ has been discussed at length in
\cite{hung2}, therefore we will refer to it for the details.)
\begin{enumerate}\item  For $\Lambda_Z \leq E \leq M_Z\,$, the coefficients
$b^{0}_Z$ and $b^{1}_Z$, appearing in Eq. (\ref{2loop}), read
explicitly \bes\label{b2l}
\be    b^0_Z=\frac{22}{3}-\frac{8}{3}\,n_{\psi}\,,\ee
\be b^1_Z=\frac{4}{3}\, (34-32\,n_{\psi})\,,\ee
 \ees
where $n_{\psi}$ denotes the number of fermions $\psi_i^{(Z)}$ (no
messenger fields are present at this stage). Choosing the mass of
the fermions to be $m_1=50\gev$ and $m_2=100\gev$, we can identify
two sub-regions: \begin{enumerate} \item $\Lambda_Z\leq E\leq
m_1$, with $n_{\psi}=0$,
\item $m_1\leq E\leq M_Z$, with $n_{\psi}=1$.
\end{enumerate}
\item Between $M_Z$ and $M_{3L}\,$ the symmetry is \be SU(2)_Z \otimes SU(3)_c \otimes
SU(2)_L \otimes U(1)_{Y}\ee and the particle content, along with
the pattern of thresholds, becomes richer. At this stage we have:
\begin{itemize}
\item the SM and the mirror fermions (the latter with a mass scale
assumed to be $M_M\sim 250 \gev$ and the number of families
$n_{MF}=3$),
\item $n_{\psi}=1,2\,$ fermionic fields $\psi_{R,L}^{(Z)}\sim(3,1,1,0)_{R,L}\subset\textbf{78}$,
\item one messenger field  $\tilde{\varphi}_1^{(Z)}\sim(3,1,2,-1/2)
\subset(3,1,3,-1/\sqrt{3})\subset(3,\overline{15})\subset\overline{\textbf{351}}$,
with a mass $M_{\tilde{\varphi}}\sim M_M\sim 250 \gev$,
\item two electro-weak Higgs doublets
\begin{itemize} \item $\varphi_1\sim(1,1,2,1/2)
\subset (1,1,\bar{3},1/\sqrt{3})\subset
(1,15)\subset\phi(\textbf{27})$,
\item $\varphi_2\sim(1,1,2,1/2)\subset (1,1,8,0)\subset
(1,35)\subset\phi(\textbf{78})$,
\end{itemize}
which we assume to
enter the renormalization group
equations at the scales $\Lambda_{\varphi_{1,2}}\sim 150\div500\gev$.
\end{itemize}
The coefficient $b_i$ can be put in the general form \bes
\be b^0_Z=\frac{22}{3}-\frac{8}{3}\,n_{\psi}-\frac{4}{3}\,n_{\tilde{\varphi}^{(Z)}}\,,  \ee
\be b^1_Z=\frac{4}{3}\, (34-32n_{\psi}-28
n_{{\tilde{\varphi}}^{(Z)}})\,,\ee
 \ees
and \bes \be b_3=11-\frac{4}{3}\,n_{F}-\frac{4}{3}\,n_{MF}\,,\ee
\be b_2=
\frac{22}{3}-\frac{4}{3}\,n_{F}-\frac{4}{3}\,n_{MF}-\frac{1}{2}
\,n_{\tilde{\varphi}^{(Z)}}-\frac{1}{6}\,n_{\varphi}\,,\ee
\be b_Y=-\frac{20}{9} \,n_{F} -\frac{20}{9}
\,n_{MF}-\frac{1}{2}\,n_{\tilde{\varphi}^{(Z)}}-\frac{1}{6}\,
n_{\varphi}\,.\ee\ees
Again, we can identify several sub-regions. Choosing for example
$\Lambda_{\varphi_{1,2}} > M_M$, we can summarize them in Table
\ref{Tableregions}.
\begin{table}
  \centering
\be \nonumber
\begin{array}{ccccccc} \hline \hline
\textrm{Energy} & & n_{F} &n_{MF} & n_{\psi} &n_{\tilde{\varphi}^{(Z)}} & n_{\varphi}   \\
  \hline
  &  & & & & & \\
M_Z\leq E\leq m_2   &  & 2 & 0 & 1 & 0 & 0 \\
m_2\leq E\leq m_t\sim 175 \gev &  & 2 &  0 & 2 & 0 & 0 \\
 m_t\leq E\leq M_M &  & 3  & 0 & 2 & 0 & 0  \\
 M_M\leq E\leq \Lambda_{\varphi_{1,2}} &  & 3 & 3 & 2 & 1 & 0  \\
\Lambda_{\varphi_{1,2}} \leq E\leq M_{3L}   &  & 3 & 3 & 2 & 1 & 2 \\
  &  & & & &  &\\
  \hline\hline
\end{array}\ee
  \caption{Regions of energy  between
the mass scales $M_Z$ and $M_{3L}$. The
parameters $n_i$ specify the number of
 particles of different type, present in each
interval.}
\label{Tableregions}
\end{table}

\item Between $M_{3L}$ and $ M_{6}\,$ the symmetry becomes
\be SU(2)_Z\otimes SU(3)_c \otimes SU(3)_L \otimes U(1)_6 \,.\ee
The fields which play a role in the evolution of the couplings at
this stage are \begin{itemize}
\item the SM, the mirror fermions and the $\psi_{R,L}^{(Z)}$
fermions introduced before,
\item  the messenger  $\varphi_1^{(Z)}\sim
 (3,1,3,-1/\sqrt{3})\subset(3,\overline{15})\subset\overline{\textbf{351}}$,
\item two electro-weak Higgs scalars, transforming, respectively, as
\begin{itemize}
\item $\varphi_1\sim(1,1,\overline{3},1/\sqrt{3})\subset (1,15)\subset \phi(\textbf{27})$,
\item $\varphi_2\sim(1,1,8,0)\subset(1,35)\subset \phi( \textbf{78})$,
\end{itemize}
\item the Higgs field, breaking $SU(3)_L\otimes U(1)_6\to SU(2)_L\otimes
U(1)_Y$, namely $\phi_{3L}\sim (1,1,6,1/\sqrt{3})\subset
(1,21)\subset\textbf{351}$.
\end{itemize}
All these particles affect the coefficients $b_i$ in the following
way: \be
    b^0_Z=0 \,,\hspace{1cm} b^1_{Z}=-96\,,\ee
 and \be b_{3}=3\,, \hspace{1cm}
b_{3L}=\frac{1}{2}\,,\hspace{1cm} b_{1,6}=-10\,. \ee

Besides them, however,  several other fields, which are part of
the $E_6$ representations listed at the beginning of the
subsection, may have a mass belonging to this region and may
therefore affect the renormalization group equations. 
For definiteness, we wil consider
\begin{itemize} \item
some fermionic multiplets, originally part of the $E_6$
representations \be \textbf{78}_{R,L}\supset (1,35)_{R,L}+\dots\ee
\item some scalar multiplets, coming from the representations
\bes  \be \textbf{351}\supset
(3,1)+(1,21)+(1,105)+\dots \,,\ee
\be \overline{\textbf{351}}\supset
(\overline{3},1)+(1,\overline{21})+(1,\overline{105})+\dots\ee\ees
\end{itemize}
and collect them in  Table \ref{TableSU6}, along with their
contribution to the coefficients $b_i$. (In the table, we show only
the representation $\textbf{351}$, $\overline{\textbf{351}}$ being
just its conjugate.)
\begin{table}
  \centering
\be \nonumber
\begin{array}{cccccccc} \hline \hline
G_1 & & G_2 & &b_{Z} & b_{3} & b_{3L} & b_{1,6} \\
  \hline
   \textrm{fermions}\,\subset \textbf{78}  &  & &  &  &  &  &  \\
   &  & (1,1,8,0) & & 0 & 0 & -2 & \;0 \\
 &  & (1,8,1,0) &  & 0 & -2 & \;0 & \;0 \\
 (1,35) &  & (1,1,1,0) & & 0 & \;0 & \;0 & \;0 \\
  &  & (1,3,\bar{3},-\frac{1}{\sqrt{3}}) & & 0 & -1 & -1 & -2 \\
   &  & (1,\bar{3},3,\frac{1}{\sqrt{3}}) & & 0 & -1 & -1 & -2 \\
\vdots  &  & & & & & &\\
 \hline
 \textrm{scalars}\, \subset\textbf{351}  &  & & &  &  &  &  \\
 &  & (3,\bar{3},1,-\frac{1}{\sqrt{3}}) & & -2 & -\frac{1}{2} & \;0 & -1 \\
 (3,15) &  & (3,1,\bar{3},\frac{1}{\sqrt{3}})  & & -2 & \;0 &
-\frac{1}{2} & -1 \\
   &  & (3,3,3,0) & & -6 & -\frac{3}{2} &-\frac{3}{2} & \;0 \\
&  & & & & & & \\
&  & (1,6,1,-\frac{1}{\sqrt{3}}) & & 0 & -\frac{5}{6} & \;0 & -\frac{2}{3} \\
 (1,21) &  & (1,1,6,\frac{1}{\sqrt{3}})  & & 0 & \;0 &
-\frac{5}{6} & -\frac{2}{3}  \\
   &  & (1,3,3,0) & & 0 & -\frac{1}{2} &-\frac{1}{2} & \;0 \\
 &  & & & & & & \\
  &  & (1,3,3,0) & & 0 & -\frac{1}{2} &-\frac{1}{2} & \;0 \\
 (1,105) &  &  (1,\bar{6},3,0)&  & 0& -\frac{5}{2} & -1 &
\;0  \\
   &  & (1,3,\bar{6},0) & & 0 & -1 &-\frac{5}{2} & \;0 \\
 \vdots & &\vdots & & & & &\\
\hline \hline
\end{array}\ee
  \caption{Values assumed by the beta function
coefficients
$b_i$, in the interval $M_{3L}\leq E\leq M_6$, for some
selected representations (fermions and scalars are
weighted in a different way, consistently with Eq.
(\ref{b})).
$G_{1,2}$ denote respectively
$G_1=SU(2)_Z \otimes SU(6)$ and
$G_2=SU(2)_Z\otimes SU(3)_c \otimes SU(3)_L\otimes U(1)_6$
 }\label{TableSU6}
\end{table}

In our analysis, we have selected
\begin{itemize}
\item among the fermions, $n_{\psi}=2$ multiplets  of $(1,1,8,0)_{R,L}$ and
$(1,8,1,0)_{R,L}\,$,
\item among the scalars, one representation $(1,1,\bar{6},1/\sqrt{3})$
(coming from a messenger $\overline{\textbf{351}}$) and two
$(1,3,3,0)$ (either coming from a $351-$dimensional representation
of Higgs or messengers).
\end{itemize}
This choice does not affect the coefficients $b_Z^{0,1}$, but modify
the others according to \be b_{3}=-6\,, \hspace{0.6cm}
b_{3L}=-\frac{28}{3}\,,\hspace{0.6cm} b_{1,6}=-\frac{32}{3}\,. \ee

\item Between $M_{6}$ and $M_{GUT}\,$, the symmetry groups are
\be SU(2)_Z \otimes SU(6)\,.\ee
The particles involved at this stage
are
\begin{itemize}
\item three families of SM and  mirror fermions, transforming jointly as
 the representations $(1,15)_{L,R}\subset\textbf{27}_{L,R}\,$;
\item $n_{\psi}=2$  fermions $\psi_{L,R}^{(Z)}\sim(3,1)_{L,R}
\subset\textbf{78}_{L,R}\,$;
\item the Higgs scalar breaking $SU(6)$ down to its subgroups, namely
$\phi_6\sim(1,35)\subset\textbf{78}\,$;
\item the two messenger fields $\tilde{\varphi}_{1,2}^{(Z)}
\sim(3,\overline{15})\subset\overline{\textbf{351}}\,$.
\end{itemize}
With this particle content, the coefficients $b_i$ read \be
b_{Z}=-18 \hspace{0.6cm}\textrm{and}\hspace{0.6cm} b_{6}=8\,. \ee As
discussed before, some other particles may acquire mass in this
region, either scalars, whose mass can take any value in the
interval of energies, or fermions, whose mass is constrained by the
strength of their Yukawa interaction with the Higgs field $\phi_6$
and can be larger than  $M_6$ within a factor ten. Again, we collect
some of them and their effect on the coefficients $b_i$ in  Table
\ref{TableE6}.
\begin{table}
  \centering
\be \nonumber
\begin{array}{cccccc} \hline \hline
E_6 & & G_1  &  & b_{Z} & b_{6} \\
  \hline
   \textrm{fermions} &  & &  &  &  \\
  & & (1,35) & & 0 & -4 \\
 \textbf{78} &  & (2,20) & & -\frac{20}{3} & -4 \\
  &  & & & & \\
 \textbf{27}  &  &(2,\bar{6}) & & -3 & -1  \\
  &  & & & & \\
 \hline
\textrm{scalars} & & & & & \\
& & (3,15) & & -10 & -2 \\
  & & (1,21)  & & 0 &  -\frac{4}{3}   \\
 \textbf{351} &  & (2,6) & & -2 & -\frac{1}{3}  \\
  & & (1,105) & & 0 & -\frac{26}{3}\\
 &  &  (2,\overline{84}) & & -14 & -\frac{38}{2}  \\
 \vdots &  &\vdots & & & \\
\hline \hline
\end{array}\ee
  \caption{Values assumed by the coefficients $b_i$
in the interval of energies $M_6\leq E\leq M_{GUT}$,
for some selected representations. Fermions and scalars
are weighetd according to Eq. (\ref{b}).
$G_1$ stands for $SU(2)_Z \otimes SU(6)$.}\label{TableE6}
\end{table}

In our analysis, we study the interplay of
\begin{itemize}
\item an additional representation  $(3,15)\subset \textbf{351}$,
entering at $M_6$ and coming from the same $\textbf{351}$ containing
the Higgs field $\phi_{3L}$ and
\item different fermionic fields, belonging to the representations
$\textbf{78}_{L,R}$ and $\textbf{27}_{L,R}$, entering at a higher
mass scale $M_{Yuk}>M_6$.
\end{itemize}
The resulting coefficients $b_i$ assume the general form  \bes
\be b_{Z}=-18-10\,n_{(3,15)}-\frac{40}{3}\,n_{(2,20)}-6\,n_{(2,\bar{6})}\,\ee
\be b_6=
8-2\,n_{(3,15)}-8\,n_{(1,35)}-8\,n_{(2,20)}-2\,n_{(2,\bar{6})}\,.\ee
\ees

\end{enumerate}

\subsection{Results}

The next step consists in evolving the coupling constants through
the different regions we have identified, taking care of the
threshold effects. Our goal is to derive the values of
\begin{itemize}
\item $M_6$ and $\alpha_6^{-1}$, where the SM groups merge into
$SU(6)$ and
\item $M_{GUT}$ and $\alpha_{GUT}^{-1}$, where $SU(2)_Z$ and $SU(6)$
unify into $E_6$.
\end{itemize}
Some parameters of our theory are free (or can vary in some definite
range), and can lead to slightly different scenarios. However, we
are not interested in studying all the possible cases in depth, but
just focus on a couple of them and show their main features.
Therefore, let us start by choosing the mass scales of the two
electro-weak Higgs doublets (discussed at point 2. of the previous
section) to assume the  values $\Lambda_{\varphi_{1,2}}=300\gev$
and $\Lambda_{\varphi_{1,2}}=150\gev$.

\begin{figure}
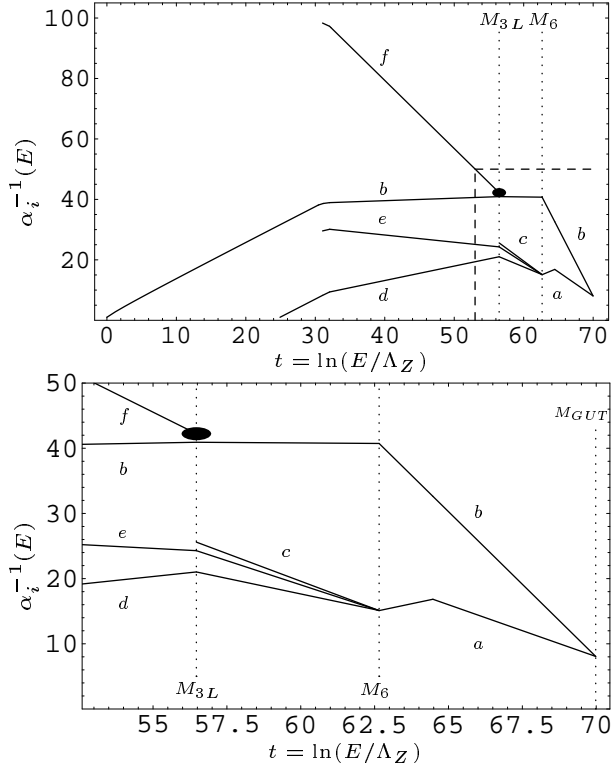

\includegraphics[angle=0,width=8cm]{GUT_mH300_2.epsi}
\includegraphics[angle=0,width=8cm]{GUT_mH300_2in.epsi}
\caption{\label{mH300_2}$\alpha_i(E)^{-1}$  versus
$t=\ln (E/\Lambda_Z)$ for $\Lambda_{\varphi_{1,2}}
= 300\,GeV$ and $0<E<M_{Planck}$ and for a restricted
range of energies.
Heavy particles entering above $M_6$ include the fermions
$ (1,35)\subset \textbf{78}$ and the scalar components
$ (3,15)\subset\textbf{351}$.
 The different indices denote:
$a=SU(6)$, $b=SU(2)_Z$,
 $c=U(1)_6$, $d=SU(3)_c$,
$e=SU(3)_L$ for $M_{3L}<E<M_6$ and $SU(2)_L$
for $E<M_{3L}$, and $f=U(1)_Y$. The black bubble stands for the
matching condition (\ref{gM3L}) at the threshold $M_{3L}$.
The mass scales read explicitly  $M_{3L}=10^{13}\gev$,
$M_6=4.75 \times 10^{15}\gev$ and $M_{GUT}=7.4 \times 10^{18}\gev$.}
\end{figure}
\begin{figure}
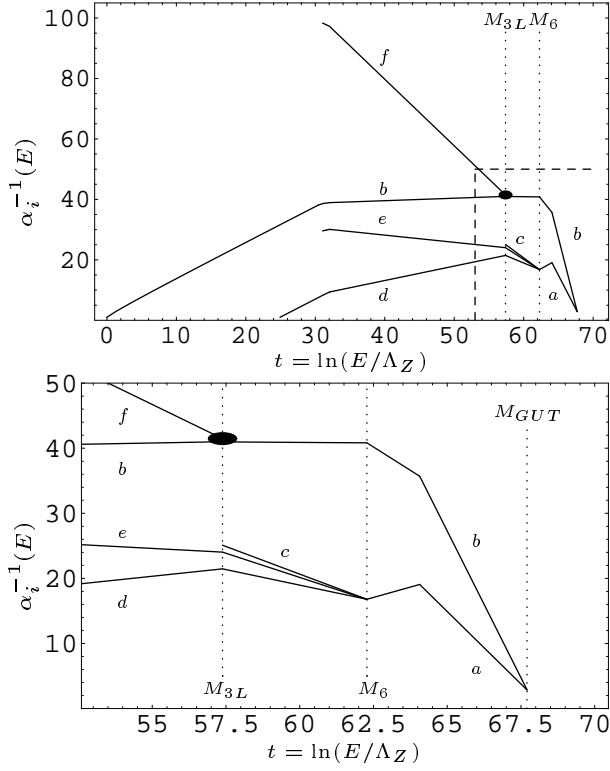

\includegraphics[angle=0,width=8cm]{GUT_mH150_1.epsi}
\includegraphics[angle=0,width=8cm]{GUT_mH150_1in.epsi}
\caption{\label{mH150_1}$\alpha_i(E)^{-1}$  versus
$t=\ln (E/\Lambda_Z)$ for $\Lambda_{\varphi_{1,2}}
= 150\,GeV$ and $0<E<M_{Planck}$ and for a restricted
range of energies.
Heavy particles entering above $M_6$ include the fermions
$ (1,35)+(2,20)\subset \textbf{78}$ and
$(2,\bar{6}) \subset\textbf{27}$.
 The different indices denote:
$a=SU(6)$, $b=SU(2)_Z$,
 $c=U(1)_6$, $d=SU(3)_c$,
$e=SU(3)_L$ for $M_{3L}<E<M_6$ and $SU(2)_L$
for $E<M_{3L}$, and $f=U(1)_Y$.
The black bubble stands for the
matching condition (\ref{gM3L}) at the threshold $M_{3L}$.
The mass scales read explicitly  $M_{3L}=2.5 \times 10^{13}\gev$,
$M_6=3.3 \times 10^{15}\gev$ and $M_{GUT}=7.6 \times 10^{17}\gev$.
}
\end{figure}

\begin{enumerate}
\item $\Lambda_{\varphi_{1,2}}=300\gev$.
\begin{itemize}
\item
With this choice and selecting the mass scale $M_{3L}\sim
10^{13}\gev$,  the unification scale $M_6$ turns out to range
approximately between the values $2.8 \times 10^{15} \gev$ and $7.2
\times 10^{15}\gev$. The uncertainty takes into account threshold's
effects and corresponds to a variation of the couplings
$\Delta{\alpha}/\alpha\equiv(\alpha_{\textrm{larger}}-\alpha_{\textrm{smaller}})/\alpha_{\textrm{smaller}}\approx
4 \% \div 4.5 \%$.

For example,  at $M_6=2.8 \times 10^{15}\gev$, we find
 $\,\alpha_{3}^{-1}=15.315$ and
$\alpha_{3L}^{-1}=16.004$  $\alpha_{16}^{-1}=15.441$, and the error
becomes $\Delta{\alpha}/\alpha\approx 4.5 \% $. Analogously, at
$M_6=7.2 \times 10^{15}\gev$ we obtain $\Delta{\alpha}/\alpha\approx
4 \% $. In Fig. \ref{mH300_2},\ref{mH300_3} (see the inset),
we show the crossing of the curves occurring at
$M_6= 4.75 \times 10^{15}\gev$, where $\alpha_{3}^{-1}=15.146$,
$\alpha_{3L}^{-1}=15.097\,$ and $\alpha_{16}^{-1}=15.137\,$.
Therefore, we take the $SU(6)$ coupling to be
\ba  && \alpha_6^{-1}\equiv\alpha(M_6)^{-1}=15.137 \hspace{0.6cm}
\textrm{at} \nonumber \\
&&  M_6= 4.75 \times 10^{15}\gev \nonumber
\ea and the corresponding
error $\Delta{\alpha_6}/\alpha_6\sim 0.3\%$.
It is straightforward
to estimate the proton partial mean lifetime associated to
these values. As
represented by $\tau_{p\to e^{+}\pi^{0}}$, it is predicted to be
\cite{PQSU(5)}
\be \label{taup}
\tau_{p\to e^{+}\pi^{0}}\approx 3.28 \times 10^{34}\left(
\frac{M_6}{3.48 \times
10^{15}}\right)^4 \left(\frac{\alpha_6^{-1}}{36.63}
\right)^2 .\ee
  Therefore,
we obtain $\tau_{p\to
e^{+}\pi^{0}}\approx 2 \times 10^{34} \textrm{yr}$, a value which
is larger than the actual lower limit of $1.47 \times
10^{32}\textrm{yr}$ \cite{Pdg}. Corrections to the central
value we
have just found come from the lower and upper bounds on the $M_6$
mass range and read, respectively, $\tau_{p\to e^{+}\pi^{0}}\approx
3 \times 10^{33} \textrm{yr}$ and $\tau_{p\to
e^{+}\pi^{0}}\approx 9 \times 10^{34} \textrm{yr}$.

\item Next, we study the interplay of different particles
above the energy threshold $M_6$ and see how they affect the
grand-unification scale $M_{GUT}$. For definiteness, we assume the
mass threshold, at which the heavy fermions acquire mass, of the
order  $\ord( 3 \times 10^{16}\gev)$. The results are collected in
Table \ref{TableGUT300} and some of them are displayed in Fig.
\ref{mH300_2},\ref{mH300_3}.

\begin{table*}
  \centering
  \be \nonumber
\begin{array}{ccccc} \hline \hline
 \textrm{fermions}& & n_{(3,15)}=0   &  &  n_{(3,15)}=1 \\
  \hline
(1,35) \subset\textbf{78}& & \textrm{no crossing} &
&\small{
\begin{array}{ll}
M_{GUT}=7.4 \times 10^{18}\gev  & \\
\alpha_{GUT}^{-1}=8.06 & \\
(\textrm{Fig.}\ref{mH300_2}) &
\end{array}
}
 \\
 &  & & &    \\
(1,35)+(2,20)\subset\textbf{78} & & \textrm{no crossing}  &
& \small{
\begin{array}{ll}
M_{GUT}=9.6 \times 10^{17}\gev  & \\
\alpha_{GUT}^{-1}=2.51 & \\
(\textrm{Fig.}\ref{mH300_3}) &
\end{array}
} \\
& & &  &    \\
\small{\left\{\begin{array}{ll}
 (1,35)+(2,20)\subset\textbf{78}  & \\
(2,\bar{6}) \subset\textbf{27} &
\end{array}
\right.}  & &\small{
\begin{array}{ll}
M_{GUT}=1.5 \times 10^{18}\gev  & \\
\alpha_{GUT}^{-1}=0.30 &
\end{array}
 } & & \small{
\begin{array}{ll}
M_{GUT}=4.5 \times 10^{17}\gev
& \\
\alpha_{GUT}^{-1}=3.92 &
\end{array}
 }\\
  & & & & \\
 \hline \hline
\end{array}\ee
 \caption{Values assumed by $M_{GUT}$ and $\alpha_{GUT}^{-1}$,
for $\Lambda_{\varphi_{1,2}}=300 \gev$ and
in the presence of different fermionic fields.
Two different scenarios are showed, corresponding to
 two ($n_{(3,15)}=0$)
and  three  ($n_{(3,15)}=1$) scalar representations transforming
as $(3,15)$ under the symmetry groups $SU(2)_Z\otimes SU(6)$.
For completeness, we quote all the values obtained from the
numerical analysis, but we restrict only  to
$\alpha_{GUT}^{-1}\gtrsim 1$
 for our physical discussion.}
\label{TableGUT300}
\end{table*}
\end{itemize}

\item $\Lambda_{\varphi_{1,2}}=150\gev $.

\begin{itemize}
\item Choosing this value for the electro-weak Higgs doublets' mass,
the mass scale $M_{3L}$ shifts to the value $2.5\times 10^{13}\gev $
and we obtain \ba && M_6=3.3\times
10^{15}\gev\hspace{0.6cm}\textrm{and}\nonumber\\
&& \alpha_6^{-1}=16.787\,,\ea
with an uncertainty due to threshold effects given by
$\Delta{\alpha_6}/\alpha_6\sim 1\%$. More generally, $M_6$ can vary
in the interval between $1.4 \times 10^{15}\gev$ and $5.9 \times
10^{15}\gev$, corresponding to
$\Delta{\alpha_6}/\alpha_6\sim 5\div 6\%$.
Using formula (\ref{taup}),
the proton partial mean lifetime reads, in this case,
 $\tau_{p\to
e^{+}\pi^{0}}\approx 6 \times 10^{33} \textrm{yr}$.
Corrections to the central
value, corresponding respectively to the lower and upper
bounds on $M_6$, are given by  $\tau_{p\to e^{+}\pi^{0}}\approx
2 \times 10^{32} \textrm{yr}$ and $\tau_{p\to
e^{+}\pi^{0}}\approx 5 \times 10^{34} \textrm{yr}$.

\item A similar analysis, performed in the range of energies between
$M_6$ and $M_{GUT}$, with a slight change in the mass scale at which
the heavy fermionic components enter, i.e. $\ord(2\times 10^{16}\gev)$,
leads to the results summarized in Table \ref{TableGUT150} and
showed in Fig. \ref{mH150_1},\ref{mH150_2},\ref{mH150_3}.

\begin{table*}
  \centering
  \be \nonumber
\begin{array}{ccccc} \hline \hline
 \textrm{fermions}& & n_{(3,15)}=0   &  &  n_{(3,15)}=1 \\
  \hline
(1,35) \subset\textbf{78}& & \textrm{no crossing} &
&\small{
\begin{array}{ll}
M_{GUT}=3.0 \times 10^{18}\gev  & \\
\alpha_{GUT}^{-1}=10.52 & \\
(\textrm{Fig.}\ref{mH150_2}) &
\end{array}
}
 \\
 &  & &  &    \\
(1,35)+(2,20)\subset\textbf{78} &  & \small{
\begin{array}{ll}
M_{GUT}=2.9 \times 10^{18}\gev  & \\
\alpha_{GUT}^{-1}=0.36 &
\end{array}
}  &
& \small{
\begin{array}{ll}
M_{GUT}=4.6 \times 10^{17}\gev  & \\
\alpha_{GUT}^{-1}=5.48 & \\
(\textrm{Fig.}\ref{mH150_3}) &
\end{array}
} \\
&  & & &    \\
\small{\left\{\begin{array}{ll}
 (1,35)+(2,20)\subset\textbf{78}  & \\
(2,\bar{6}) \subset\textbf{27} &
\end{array}
\right.}  & &\small{
\begin{array}{ll}
M_{GUT}=7.6 \times 10^{17}\gev  & \\
\alpha_{GUT}^{-1}=2.82 & \\
(\textrm{Fig.}\ref{mH150_1}) &
\end{array}
 } & & \small{
\begin{array}{ll}
M_{GUT}=2.3 \times 10^{17} \gev
& \\
\alpha_{GUT}^{-1}=6.75 &
\end{array}
 }\\
  & & & & \\
 \hline \hline
\end{array}\ee
 \caption{Values assumed by $M_{GUT}$ and $\alpha_{GUT}^{-1}$,
for $\Lambda_{\varphi_{1,2}}=150 \gev$ and
in the presence of different fermionic fields.
Two different scenarios are showed, corresponding to
 two ($n_{(3,15)}=0$)
and  three  ($n_{(3,15)}=1$) scalar representations transforming
as $(3,15)$ under the symmetry groups $SU(2)_Z\otimes SU(6)$.}
\label{TableGUT150}
\end{table*}
\end{itemize}
\end{enumerate}

\section{Oblique corrections}

In this section we analyze the oblique corrections due to the
introduction of new extra particles in the spectrum, below the
$\tev$ scale.
They are expressed in terms of three
parameters, $S$, $T$ and $U$  \cite{STU}, whose values can
be extracted from the
electro-weak precision measurement data \cite{Pdg}.
The new (with respect to the SM reference point)
 contributions to $S$, $T$ and $U$  come from:
\bi \item  the three extra families of chiral mirror fermions,
\item the two Higgs doublets (see Section \ref{fmass}),
\item the messenger field $\tilde{\varphi}^{(Z)}_1\sim
(3,1,2, Y_{\tilde{\varphi}}=-1)$,
\ei and are additive.

We notice that the interplay between extra families of chiral
fermions, whose masses range  from $\ord(50 \gev)$ to  the
electro-weak scale, and the Higgs sector, spanning a certain
interval of masses,
 have already been discussed by
He \textit{et al} \cite{Polonsky}. In particular, they have shown
that \textit{three} mirror families would be inconsistent with the
electro-weak precision data, unless \textit{two} Higgs doublets were
present. This  result applies straightforwardly to our situation,
which displays a choice of masses (both for the fermions and the for
the Higgs) compatible with  the analysis performed in
\cite{Polonsky}.

\begin{figure}
\includegraphics[angle=0,width=8cm]{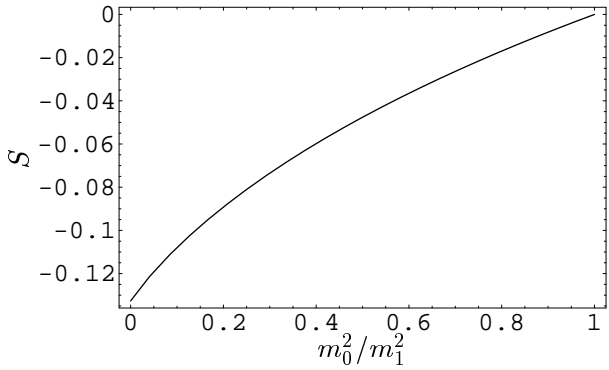}
\caption{\label{Smess} Correction to the $S$ parameter produced by
the messenger field $\tilde{\varphi}^{(Z)}$, where
 $m_{0,1}$ denote, respectively, the mass of the custodial
representations $J=0$ and $J=1$.
$S$ ranges in the interval
$-0.13\lesssim S \lesssim 0$, where $S=0$ corresponds to no
splitting between the masses of the custodial representations
$J=0$ and $J=1$, and $S=-0.13$
refers to the limit $\frac{m_0^2}{m_1^2}\to 0$.}
\end{figure}

What remains to be proven is that also the oblique corrections to
$S$ and $T$, due to the messenger field $\tilde{\varphi}^{(Z)}_1$,
can be very small. Assuming the three $SU(2)_L$ doublets to be
quasi-degenerate, the corresponding contribution to $T$ is indeed
negligible. The correction to $S$ can be evaluated  by noting that
the custodial isospin symmetry is not violated \cite{S}. Assigning a
doublet of messenger fields to the representation
$(j_L,j_R)=(1/2,1/2)$ of the global symmetry group $SU(2)_L\otimes
SU(2)_R\otimes U(1)_Y$, the correction to $S$ can be visualized in
Fig. \ref{Smess}. The value of $S$ ranges in the interval
$-0.13\lesssim S \lesssim 0$, where $S=0$ corresponds to no
splitting between the masses of the custodial representations $J=0$
and $J=1$, i.e. $m_0^2/m_1^2=1$, and $S=-0.13$ refers to the limit
$m_0^2/m_1^2\to 0$.

In summary, the contributions to the $S$, $T$ and $U$ parameters
from new particles contained in our model are compatible with the
data.


\section{Conclusion}

We have presented, in this paper, a grand unified model based on the
group $E_6 \supset SU(2)_Z \otimes SU(6)$. The main rationale for
this unification is the presence of an {\em unbroken}
gauge group $SU(2)_Z$ which
forms the cornerstone of the dark energy model of \cite{hung1},
\cite{hung2}. In that model, the $SU(2)_Z$ gauge coupling was assumed
to have an initial value at high energies ($\sim 10^{16}\,GeV$)
of the same order as the SM couplings at comparable energies
and grows strong i.e. $\alpha_Z \equiv g_Z^2/4\,\pi\sim O(1)$ at a scale
$\Lambda_Z \sim 3 \times 10^{-3}\,eV$. In that analysis, the masses
of the fermions $\psi_{i}^{(Z)}$ ($i=1,2$) were found to be
important in the evolution of $\alpha_Z$ with a value
$\sim 200\,GeV$ found to be particularly relevant for these
particles to be CDM candidates.
Around $\Lambda_Z$, $SU(2)_Z$
instantons induce a potential for an axion-like particle $a_Z$ which
plays the role of the acceleron in that model.

The main purpose
of the present paper is to show that, with the desired values of the
parameters presented in \cite{hung2}, one can indeed find a scenario
in which $SU(2)_Z$ is unified with the SM in a way that was
described above. In a nutshell, we found that $SU(2)_Z$ can be
embedded in $E_6$ via the route $E_6 \rightarrow SU(2)_Z \otimes SU(6)$.
As we have discussed in the previous sections, one of the
symmetry breaking route that we choose to present here for $SU(6)$
is $SU(6) \rightarrow SU(3)_c \otimes SU(3)_L \otimes U(1)_6
\rightarrow SU(3)_c \otimes SU(2)_L \otimes U(1)_Y$.
Some of the highlights of this unification scenario are the following:

1) The presence of heavy right-handed mirror quarks and leptons which
could be produced and searched for at future colliders such as the
Large Hadron Collider (LHC): the lightest among them can decay into
a SM fermion and $W$ through a mixing between the
$SU(3)_L/SU(2)_L$ gauge bosons with $W$'s, or it can decay
into a SM fermion and $\phi_S(\textbf{1})$ if the latter
is light enough. 
Finally, as discussed in \cite{hung2}, \cite{hung3}, the
so-called ``progenitor of SM lepton numbers'', the
messenger scalar field ${\tilde{\bm{\varphi}}}^{(Z)}$,
can be produced at the LHC e.g. via electroweak gauge boson fusion
processes (since it does not carry color), and subsequently
decays into a SM lepton and $\psi_{i}^{(Z)}$ with interesting
signatures. The full discussion is beyond the scope of this paper.

2) The SM gauge couplings eventually merge (through various steps)
into the $SU(6)$ coupling at $\sim 5 \times 10^{15} \,GeV$. The
value of $SU(6)$-GUT coupling is generally higher than a
typical GUT value (even for supersymmetric GUT): $\alpha_{6}^{-1}
(M_6) \sim 20$ as opposed to a non-supersymmetric GUT value
$\sim 40$. Since one expects the lifetime of the proton
to be proportional to $M_{6}^{4}/\alpha_{6}^2$, it is 
shown in Section (\ref{RG})
that the estimated proton lifetime is well above the current lower bound;

3) The $E_6$ GUT scale, $M_{GUT}$ is shown in Section (\ref{RG}) to be
at least two orders of magnitude higher than $M_6$. We found that
typically $\alpha_{GUT}^{-1} \alt 10$. It is interesting to note
that the $E_6$ coupling at the unification scale $M_{GUT}$ is not
too far from the strong coupling regime $\alpha_{GUT} \sim 1$;

4) As shown in Section (\ref{fmass}), the masses of the ``up'' and ``down''
fermion sectors automatically come from couplings to
two {\em different} Higgs sectors.

\begin{acknowledgments}
This work is supported
in parts by the US Department of Energy under grant No.
DE-A505-89ER40518. PQH would like to thank James(bj) Bjorken
for the probing question concerning the possibility
of embedding the dark energy model into the $E_6$ group.
We would like to thank Lia Pancheri, Gino Isidori
and the Spring Institute for the hospitality in the Theory Group
at LNF, Frascati, where part of this work was carried out.
\end{acknowledgments}

\begin{figure}
\includegraphics[angle=0,width=8cm]{GUT_mH300_3.epsi}
\includegraphics[angle=0,width=8cm]{GUT_mH300_3in.epsi}
\caption{\label{mH300_3}$\alpha_i(E)^{-1}$  versus
$t=\ln (E/\Lambda_Z)$ for $\Lambda_{\varphi_{1,2}}
= 300\,GeV$ and $0<E<M_{Planck}$ and for a restricted
range of energies.
Heavy particles entering above $M_6$ include the fermions
$ (1,35)+(2,20)\subset \textbf{78}$ and the scalar components
$ (3,15)\subset\textbf{351}$.
 The different indices denote:
$a=SU(6)$, $b=SU(2)_Z$,
 $c=U(1)_6$, $d=SU(3)_c$,
$e=SU(3)_L$ for $M_{3L}<E<M_6$ and $SU(2)_L$
for $E<M_{3L}$, and $f=U(1)_Y$.
 The black bubble stands for the
matching condition (\ref{gM3L}) at the threshold $M_{3L}$.
The mass scales read explicitly  $M_{3L}=10^{13}\gev$,
$M_6=4.75 \times 10^{15}\gev$ and $M_{GUT}=9.6 \times 10^{17}\gev$.
}
\end{figure}

\begin{figure}
\includegraphics[angle=0,width=8cm]{GUT_mH150_2.epsi}
\includegraphics[angle=0,width=8cm]{GUT_mH150_2in.epsi}
\caption{\label{mH150_2}$\alpha_i(E)^{-1}$  versus
$t=\ln (E/\Lambda_Z)$ for $\Lambda_{\varphi_{1,2}}
= 150\,GeV$ and $0<E<M_{Planck}$ and for a restricted
range of energies.
Heavy particles entering above $M_6$ include the fermions
$ (1,35)\subset \textbf{78}$ and the scalar components
$ (3,15)\subset\textbf{351}$.
 The different indices denote:
$a=SU(6)$, $b=SU(2)_Z$,
 $c=U(1)_6$, $d=SU(3)_c$,
$e=SU(3)_L$ for $M_{3L}<E<M_6$ and $SU(2)_L$
for $E<M_{3L}$, and $f=U(1)_Y$.
The black bubble stands for the
matching condition (\ref{gM3L}) at the threshold $M_{3L}$.
The mass scales read explicitly  $M_{3L}=2.5 \times 10^{13}\gev$,
$M_6=3.3 \times 10^{15}\gev$ and $M_{GUT}=3.0 \times 10^{18}\gev$.
}
\end{figure}
\begin{figure}
\includegraphics[angle=0,width=8cm]{GUT_mH150_3.epsi}
\includegraphics[angle=0,width=8cm]{GUT_mH150_3in.epsi}
\caption{\label{mH150_3}$\alpha_i(E)^{-1}$  versus
$t=\ln (E/\Lambda_Z)$ for $\Lambda_{\varphi_{1,2}}
= 150\,GeV$ and $0<E<M_{Planck}$ and for a restricted
range of energies.
Heavy particles entering above $M_6$ include the fermions
$ (1,35)+(2,20)\subset \textbf{78}$ and the scalar components
$ (3,15)\subset\textbf{351}$.
 The different indices denote:
$a=SU(6)$, $b=SU(2)_Z$,
 $c=U(1)_6$, $d=SU(3)_c$,
$e=SU(3)_L$ for $M_{3L}<E<M_6$ and $SU(2)_L$
for $E<M_{3L}$, and $f=U(1)_Y$.
The black bubble stands for the
matching condition (\ref{gM3L}) at the threshold $M_{3L}$.
The mass scales read explicitly  $M_{3L}=2.5 \times 10^{13}\gev$,
$M_6=3.3 \times 10^{15}\gev$ and $M_{GUT}=4.6 \times 10^{17}\gev$.
}
\end{figure}

\end{document}